\documentclass[epj]{svjour}
\usepackage[dvips]{epsfig}
\usepackage{subfigure}
\usepackage{color}
\usepackage{amsmath,amssymb,bm}
\newcommand{\nc}[1]{\newcommand{#1}}
\nc{\its}[1]{\itshape #1 \upshape}
\nc{\mc}[3]{\multicolumn{#1}{#2}{#3}}
\nc{\bc}{\begin{center}}
\nc{\ec}{\end{center}}
\nc{\ig}[1]{\bc \includegraphics{#1} \ec}
\nc{\bo}[1]{\mbox{\boldmath \( #1 \! \! \)  \unboldmath}}
\nc{\beq}{\begin{equation}}
\nc{\eeq}{\end{equation}}
\nc{\bew}{\begin{eqnarray}}
\nc{\eew}{\end{eqnarray}}
\nc{\bs}{\begin{subeqnarray}}
\nc{\es}{\end{subeqnarray}}
\nc{\nnn}{\nonumber \\}
\nc{\Mvariable[1]}{\; #1}
\nc{\f}[2]{\frac{#1}{#2}}
\nc{\td}[2]{\f{d #1}{d #2}}
\nc{\pd}[2]{\f{\partial #1}{\partial #2}}
\nc{\suli}{\sum\limits}
\nc{\proli}{\prod\limits}
\nc{\ili}{\int\limits}
\nc{\sr}[2]{\stackrel{#1}{#2}}
\nc{\dps}{\displaystyle}
\nc{\ket}[1]{\left| #1 \right>}
\nc{\bra}[1]{\left< #1 \right|}
\nc{\bracket}[2]{\left< #1 \right| \left. \! #2 \right>}
\nc{\norm}[1]{\left\| #1 \right\|}
\nc{\lndm}[1]{\pd{^{#1} \ln{\det{M}}}{\mu^{#1}}}
\nc{\pdmm}[1]{M^{-1} \pd{^{#1} M}{\mu^{#1}}}
\nc{\pdm}{M^{-1}\pd{M}{\mu}}
\nc{\trac}[1]{\mbox{Tr}\left(#1\right)}
\nc{\fb}{\color{blue}}
\nc{\fr}{\color{red}}
\nc{\muh}{\hat \mu}
\nc{\nuh}{\hat \nu}
\nc{\rhoh}{\hat \rho}
\nc{\sigmah}{\hat \sigma}
\def \beq{\begin{equation}}
\def \eeq{\end{equation}}
\def \beqa{\begin{eqnarray}}
\def \eeqa{\end{eqnarray}}
\def \lt{\left}
\def \rt{\right}
\def \c{\chi}
\def \cb{\bar\chi}
\def \bx{\bm x}

\def \cM{{\mathcal M}}
\def \unitmatrix{1\!\!\!1}

\def \jhep{{JHEP\ }}
\def \npb{{Nucl.\ Phys.\ B}}
\def \npsl{{Nucl.\ Phys.\ Proc.\ Suppl.\ }}
\def \plb{{Phys.\ Lett.\ B}}
\def \prd{{Phys.\ Rev.\ D}}

\def \prl{{Phys.\ Rev.\ Lett.\ }}

\def \pos{{PoS\ }}

\def \eg{{\sl e.g.\ }}
\def \ie{{\sl i.e.\ }}
\def \viz{{\sl viz.\ }}
\def \etal{{\sl et al.}}

\begin{document}
\title{Meson screening masses from lattice QCD with two light and the strange quark}
\author{
M. Cheng\inst{1}, 
S. Datta\inst{2},
A. Francis\inst{3},
J. van der Heide\inst{3},
C. Jung\inst{4},
O. Kaczmarek\inst{3},
F. Karsch\inst{3,4},
E. Laermann\inst{3},
R. D. Mawhinney\inst{5},
C. Miao\inst{4},
S. Mukherjee\inst{4},
P. Petreczky\inst{4},
J. Rantaharju\inst{3,6},
C. Schmidt\inst{7,8}
and
W. S\"oldner\inst{7,8} \\[2pt]
}
\institute{
Physics Division, Lawrence Livermore National Laboratory, Livermore, CA 94550, USA
\and
Department of Theoretical Physics, Tata Institute of Fundamental Research, Mumbai
400005, India
\and
Fakult\"at f\"ur Physik, Universit\"at Bielefeld, D-33615 Bielefeld, Germany 
\and
Physics Department, Brookhaven National Laboratory, Upton, NY 11973, USA 
\and
Physics Department, Columbia University, New York, NY 10027, USA
\and
Department of Physics, University of Helsinki, FI-00014, Finland
\and
Frankfurt Institute for Advanced Studies, J.W.Goethe Universit\"at Frankfurt, D-60438
Frankfurt am Main, Germany
\and
GSI Helmholtzzentrum f\"ur Schwerionenforschung, D-64291 Darmstadt, Germany
}
\abstract{
We present results for screening masses of mesons built from light and strange quarks
in the temperature range of approximately between $140$ MeV to $800$ MeV. The lattice
computations were performed with $2+1$ dynamical light and strange flavors of
improved (p4) staggered fermions along a line of constant physics defined by a pion
mass of about $220$ MeV and a kaon mass of $500$ MeV. The lattices had temporal
extents $N_\tau = 4,6$ and $8$ and aspect ratios of $N_s/N_\tau\geq4$. At least up to
a temperature of $140$ MeV the pseudo-scalar screening mass remains almost equal to
the corresponding zero temperature pseudo-scalar (pole) mass. At temperatures around
$3T_c$ ($T_c$ being the transition temperature) the continuum extrapolated
pseudo-scalar screening mass approaches very close to the free continuum result of
$2\pi T$ from below. On the other hand, at high temperatures the vector screening
mass turns out to be larger than the free continuum value of $2\pi T$. The
pseudo-scalar and the vector screening masses do not become degenerate even for a
temperature as high as $4T_c$. Using these mesonic spatial correlation functions we
have also investigated the restoration of chiral symmetry and the effective
restoration of the axial symmetry. We have found that the vector and the axial-vector
screening correlators become degenerate, indicating chiral symmetry restoration, at a
temperature which is consistent with the QCD transition temperature obtained in
previous studies. On the other hand, the pseudo-scalar and the scalar screening
correlators become degenerate only at temperatures larger than $1.3T_c$, indicating
that the effective restoration of the axial symmetry takes place at a temperature
larger than the QCD transition temperature.
}
\authorrunning{ }
\titlerunning{ }
\date{\bc\today\ec}
\maketitle
\section{Introduction}          \label{se.intro}

Hadron correlations at finite temperature have been advocated since long as a means
to learn about non-perturbative properties of the quark gluon plasma from lattice
simulations \cite{detar}. Studies of the in-medium properties of hadronic excitations
provide information about some important length-scales at high temperatures and give
an idea about the relevant degrees of freedom in the plasma and their possible
physical effects. Furthermore, these studies also illuminate aspects of the chiral
and the effective anomalous $U_A(1)$ symmetry restorations in QCD. 

Most lattice analyses of hadronic excitations have concentrated on spatial
correlation functions. Different from correlators in the temporal direction the
physical extent of which is limited by the inverse temperature, a spatial correlation
can be studied, in principle, at arbitrarily large distances facilitating the
isolation of the lowest excitation. The exponential decay of these spatial
correlators defines the so-called \emph{screening masses} \cite{detar}.

Physically, the inverse of a screening mass can immediately be interpreted as the
spatial distance beyond which the effects of putting a test hadron in the medium are
effectively screened.  The screening masses ($M$) need not be identical to the masses
($m$) defined via the exponential decay of temporal correlators, but they are related
to the same spectral function. Thus, amongst other purposes, they provide a test bed
for model building of the spectral function. 

Mesonic spatial correlation functions can be very useful tools to investigate the
chiral and effective $U_A(1)$ symmetry restorations in QCD. Symmetry restorations
demand that certain spatial (and also the temporal) correlation functions become
identical at all distances. Of course, this also means that the screening masses of
these mesons become degenerate as a symmetry gets restored. However, when one is
working with a limited spatial extent, \ie with relatively smaller volumes,
extraction of screening masses from the large distance behavior of the correlators
becomes difficult. Even in such cases one can obtain indications of symmetry
restorations by studying the degeneracies between certain mesonic spatial correlation
functions themselves at relatively shorter distances. Moreover, such symmetry
restorations can be studied even without the use of the computationally demanding
quark-line disconnected diagrams. As for example, restoration of chiral symmetry can
be studied through the degeneracy of the spatial correlation functions belonging to
the vector ($\rho$) and the axial-vector ($a_1$) channels. It is known \cite{ua1}
that in the limit of mass-less light quarks the 2-point pseudo-scalar and scalar
correlation functions of the $2+1$ flavor theory are also sensitive to topologically
nontrivial configurations similar to the case of a 2 flavor theory. Hence the
effective restoration of $U_A(1)$ symmetry at high temperature can also be studied
using the degeneracy between  spatial correlation functions in the pseudo-scalar
($\pi$) and the iso-triplet scalar ($a_0$) channels. 

Accurate non-perturbative determination of mesonic screening masses also plays an
important role in testing the applicability of the dimensionally reduced perturbation
theory \cite{Appelquist}. Within the framework of dimensional reduction at a
particular temperature ($T$) one integrates out all the non-zero bosonic modes and
all the fermionic modes of the $3+1$ dimensional theory in order to obtain an
effective $3$ dimensional theory, for distance scales $RT\gg1$, consisting of static
gauge fields coupled to adjoint scalars. Such a procedure is justified only when the
spatial correlation lengths associated with all the fermionic modes in the original
$3+1$ dimensional theory are smaller than the largest spatial correlation length
associated with the bosonic modes. Thus the applicability of dimensional reduction,
at a certain temperature and distance scale, crucially depends on the assumption that
the screening masses of all the objects made out of fermionic modes of the original
$3+1$ theory are larger than the lowest screening mass associated with objects
consisting only of the original gauge fields. In fact, this issue was addressed in
Ref.\ \cite{GG-glue} in $4$ flavor QCD using the standard staggered action on lattice
with temporal extent $N_\tau=4$. This study suggests that the pseudo-scalar
quark--anti-quark (`pion') correlator gives the largest spatial correlation length
(\ie smallest screening mass) for temperatures as high as $T=3T_c$, $T_c$ being the
chiral transition temperature, indicating the breakdown of dimensional reduction
below that temperature. On the other hand, another recent study \cite{spatial-string}
in $2+1$ flavor QCD using an improved staggered fermion action found that the
non-perturbative values of the spatial string tension are in good agreement with the
prediction of dimensionally reduced perturbation theory \cite{mikko} right down to
$T=1.5T_c$. Thus in order to clarify this issue related to the applicability of
dimensional reduction it is necessary to have a detailed study of the meson screening
masses in $2+1$ flavor QCD using an improved staggered fermion action and small
lattice spacings.

In addition to the test of the applicability of the dimensional reduction framework,
improved determinations of mesonic screening masses are also important for checking
the accuracy of the prediction of re-summed perturbation theories. In mass-less free
field theory the meson screening masses acquire a value of $2\pi T$ \cite{Friman}
independent of the $J^{PC}$ structure of the interpolating meson operators. The
leading order perturbative corrections of $\mathcal{O}(g^2T)$ to this free field
value have been calculated using both the dimensional reduction framework
\cite{Mikko} and the hard thermal loop framework \cite{Alberico}. It was found that
the leading order perturbative correction is positive and equal for all the meson
channels.  Thus for large temperatures the perturbative results approach the free
field value from above in the same way independent of the meson channel.

On the lattice, screening masses have been studied in the quenched approximation and
with dynamical quarks, both in Wilson as well as staggered type fermion
discretizations (for a review and references see e.g. \cite{karsch} and references
therein). Regarding the general pattern, little differences between quenched\-  and
full QCD studies have been observed. Below the transition region differences between
masses and screening masses were found to be small, independent of the lattice
fermion action.  Above the transition mesonic screening masses were found to approach
the zero quark mass continuum free field limit of $2\pi T$ from below. In the
temperature range of 2 to 4 times the transition temperature vector screening masses
have been closer to this value while pseudo-scalar ones were found to be typically 10
\% smaller than the vectors for Wilson quarks and somewhat lower values were reported
for the standard staggered action. This is in contrast to the perturbative
predictions as mentioned above.

However, most of the studies were performed at fixed lattice spacing and fixed volume
($\mathcal{V}$). More recently, the lattice spacing dependence of screening masses
based on the standard staggered action was analyzed in \cite{contstagg}. The results
show that discretization effects are significant for this choice of action and the
pseudo-scalar screening mass at a fixed value of $T\mathcal{V}^{1/3}$ approaches the
infinite volume free lattice value from below. Preliminary results of a systematic
study of volume as well as lattice spacing dependence of the screening masses for
quenched improved Wilson fermions on large lattices have shown \cite{Wissel,me} that
in this discretization scheme the volume effects increase the screening masses and
discretization effects work in the opposite direction.  On the other hand, Ref.
\cite{overlap} concludes that finite volume as well as discretization effects are
small for quenched overlap fermions, with the vector being close to $2\pi T$ and the
scalar being $10\%$ lower at $T=2T_c$. 

Here we present lattice results for meson screening masses in $2+1$ flavor QCD using
improved p4 \cite{p4} staggered fermions with the zero temperature Goldstone pion
mass of $220$ MeV and the zero temperature kaon mass of $500$ MeV. The temperature
ranges from about $140$ MeV to maximally $820$ MeV. Our study is performed at three
different sets of the lattice spacings corresponding to $N_\tau=4,\ 6$ and $8$, which
allows to address discretization effects. 

The paper is organized such that after presenting the numerical set-up in
section~\ref{se.operators} we briefly discuss discretization and finite volume
effects as they arise, for our choice of action, in semi-analytical computations in
the free case.  This provides the background on which we present and discuss our
results on screening masses in section~\ref{se.light}. They have been determined
from spatial correlators of operators built with degenerate up ($u$) and down ($d$)
quarks as well as with strange ($s$) quarks with the same masses as in the generation
of the configurations.  Preliminary results have been presented earlier in
\cite{me,Swagato}.

\section{Numerical set-up}      \label{se.operators}

\begin{table}[!t]
\begin{center}
\begin{tabular}{c c|c c|c c|c c} \hline
           & $\phi(\bx)$& \multicolumn{2}{c|}{$\Gamma$}&
  \multicolumn{2}{c|}{$J^{PC}$}& \multicolumn{2}{c}{states} \\
  && NO& O& NO& O& NO& O \\ \hline

  $\cM1$& $(-1)^{x+y+\tau}$& $\gamma_3\gamma_5$& $\unitmatrix$& $0^{-+}$& $0^{++}$&
  $\pi_2$& $a_0$ \\

  $\cM2$& $1$& $\gamma_5$& $\gamma_3$& $0^{-+}$& $0^{+-}$& $\pi$& -- \\

  $\cM3$& $(-1)^{y+\tau}$& $\gamma_1\gamma_3$& $\gamma_1\gamma_5$& $1^{--}$& $1^{++}$&
  $\rho^\mathcal{T}_2$& $a_1^\mathcal{T}$ \\ 

  $\cM4$& $(-1)^{x+\tau}$& $\gamma_2\gamma_3$& $\gamma_2\gamma_5$& $1^{--}$& $1^{++}$&
  $\rho^\mathcal{T}_2$& $a_1^\mathcal{T}$ \\

  $\cM5$& $(-1)^{x+y}$& $\gamma_4\gamma_3$& $\gamma_4\gamma_5$& $1^{--}$& $1^{++}$&
  $\rho^\mathcal{L}_2$& $a_1^\mathcal{L}$ \\

  $\cM6$& $(-1)^x$& $\gamma_1$& $\gamma_2\gamma_4$& $1^{--}$& $1^{+-}$&
  $\rho^\mathcal{T}_1$& $b_1^\mathcal{T}$ \\

  $\cM7$& $(-1)^y$& $\gamma_2$& $\gamma_1\gamma_4$& $1^{--}$& $1^{+-}$&
  $\rho^\mathcal{T}_1$& $b_1^\mathcal{T}$ \\

  $\cM8$& $(-1)^\tau$& $\gamma_4$& $\gamma_1\gamma_2$& $1^{--}$& $1^{+-}$&
  $\rho^\mathcal{L}_1$& $b_1^\mathcal{L}$ \\ \hline 
\end{tabular}
\end{center}
\caption{The list of meson operators studied in this work. States associated with the
non-oscillating and the oscillating part of the screening correlators are designated
by the identifiers NO and O respectively. Particle assignments of the corresponding
states are given only for the $u$-$d$ flavor combination.}
\label{tb.phase_factors}
\end{table}

In the staggered lattice formulation of QCD meson operators can be defined
\cite{Bengt} as $\cM = \bar\psi\lt(\Gamma^D\otimes\Gamma^F\rt)\psi$, where the
fermion field $\psi$ has four Dirac- and four flavor/taste- components and is defined
on a coarse lattice with lattice spacing $2 a$.  The matrices $\Gamma^D$ and
$\Gamma^F$ are products of $\gamma$-matrices and generate the spin-flavor/taste
structure of the corresponding meson.  Here we are interested in local meson
operators, for which $\Gamma^D=\Gamma^F\equiv\Gamma$. In terms of staggered fermion
fields $\c(\bx)$ which are defined at the points $\bx = (x,y,z,\tau)$ of a fine
lattice with spacing $a$, these local meson operators can be written as $\cM(\bx) =
\tilde\phi(\bx) \cb(\bx)\c(\bx)$, where $\tilde\phi(\bx)$ is a phase factor depending
on the choice of $\Gamma$. 

In this work we study mesons built from $\bar ud$, $\bar us$ and $\bar ss$
flavor combinations. For the flavor non-singlet combinations $\bar ud$ and $\bar us$
the correlation function consists solely of the quark-line connected part. However,
for the flavor singlet $\bar ss$ flavor combination we only use the connected part of
the correlation function and neglect the computationally demanding quark-line
disconnected part. The connected part of the staggered meson screening correlator,
projected to zero transverse momentum, ${\bm p}_\perp = (p_x,p_y)$, and to zero
(boson) Matsubara frequency, $\omega_n$, is obtained as
\beq
C(z) = \sum_{x,y,\tau} \phi(\bx)  
\lt\langle\lt(M^{-1}_{\bm{0}{\bx}}\rt)^\dagger M^{-1}_{{\bm 0}\bx} \rt\rangle, 
\eeq
where $M^{-1}_{{\bm 0}\bx}$ is the full staggered propagator from ${\bm 0}$ to $\bx$.
Since a staggered fermion meson correlator, in general, contains two different mesons
with opposite parity \cite{golterman} the correlator is parametrized as
\bew
C(z) = && A_{NO} \cosh\lt[ M_- \lt( z - \frac{N_s}{2} \rt)\rt] \nnn 
& - & (-1)^z A_O \cosh\lt[ M_+ \lt( z - \frac{N_s}{2} \rt)\rt] .
\label{eq.cor-par}       
\eew
Different from \eg \cite{golterman,altmeyer} we have chosen the phase factor in
Eq.(\ref{eq.cor-par}) as $\phi(\bx)=-(-1)^{x+y+\tau}\tilde\phi(\bx)$.  According to
this convention $M_-$ ($M_+$) corresponds to the screening mass of the lightest
negative (positive) parity state\footnote{For the screening states we keep, as usual,
the $J^{PC}$ assignments as they would emerge, for the same operators and correlation
functions, in the zero temperature limit.} and comes as the non-oscillating, NO
(oscillating, O), part of the screening correlator.  For instance, in this convention
the Goldstone pion comes as the non-oscillating part of the screening correlator.
Note that both amplitudes are positive \cite{altmeyer}, $A_{NO}, A_O \ge 0$. 

For staggered fermions there are 8 possible local meson operators \cite{golterman}
which are listed in Table\ \ref{tb.phase_factors}, together with the corresponding
phase factors $\phi(\bx)$.  We separate between the transverse ($\mathcal{T}$) and
longitudinal ($\mathcal{L}$) polarizations of vector and axial-vector correlators.
At $T=0$ the rotational symmetry group of the lattice corresponds to the continuum
$O(3)$. For spatial correlators, however, at non-zero temperature this symmetry
breaks down to $O(2)\times Z(2)$ \cite{gupta}.  Thus, the longitudinal states are
allowed to acquire screening masses different from the transverse ones.

For the present study we have analyzed the gauge configurations generated by means of
the RHMC algorithm \cite{algo} by the RBC-Bielefeld \cite{spatial-string,RBCBi-eos}
and the HotQCD \cite{hotQCD} collaboration using the p4 staggered action. These
configurations were generated for $2$ degenerate dynamical light quarks and $1$
heavier dynamical strange quark along the Line of Constant Physics (LCP). The LCP is
obtained by tuning the bare quark masses such that at zero temperature the
(Goldstone) pion mass $m_\pi$ is approximately $220$ MeV and the kaon mass $m_K$
equals $500$ MeV.  As in \cite{RBCBi-eos}, relative scales have been obtained from
the Sommer $r_0$ parameter \cite{Rainer}, absolute ones from bottomonium splittings
\cite{Gray} resulting in $r_0=0.469$ fm in the continuum limit for physical quark
masses. For more details about the simulations and scale settings confer
Ref.\cite{RBCBi-eos}.  The values of coupling constant and bare light quark mass in
lattice units at which these configurations were generated are summarized in Table
\ref{tb.runpara}. The bare strange quark mass has always been $10$ times larger,
$m_s=10m_l$. The table also contains the temperature values and lists the number of
configurations which were used for the present analysis. The configurations were
separated by 10 trajectories.  Autocorrelation times were estimated from the
point-to-point spatial correlation functions at intermediate to large separations $z$
from source and sink.  The largest autocorrelations were observed in the lightest
pseudo-scalar channel, $\cM2$, and turned out to be of ${\cal O}(10)$ configurations,
\ie ${\cal O}(100)$ trajectories, in the vicinity of the transition and dropping
quickly away from it. 

\begin{table}[!t]
\begin{center}
\begin{tabular}{|ll|lr|lr|lr|}
\hline
        &       & \multicolumn{2}{|c|}{$16^3 \times 4$} &
                  \multicolumn{2}{|c|}{$24^3 \times 6$} &
                  \multicolumn{2}{|c|}{$32^3 \times 8$} \\
$\beta$ & $m_l$ & $T$ & conf. &
                  $T$ & conf. &
                  $T$ & conf. \\
\hline
3.290 & 0.00650 & 192 & 1600 &     &     &     &    \\
3.320 & 0.00650 & 201 & 1600 &     &     &     &    \\
3.351 & 0.00591 & 218 &  800 & 145 & 984 &     &    \\
3.382 & 0.00520 & 234 &  600 &     &     &     &    \\
3.410 & 0.00412 & 263 & 1000 & 175 & 584 &     &    \\
3.430 & 0.00370 &     &      & 186 & 514 & 139 & 783\\
3.445 & 0.00344 &     &      & 197 & 949 &     &    \\
3.455 & 0.00329 &     &      & 200 & 438 &     &    \\
3.460 & 0.00313 & 304 &  600 & 203 & 406 &     &    \\
3.490 & 0.00290 &     &      & 226 & 824 &     &    \\
3.500 & 0.00253 &     &      &     &     & 175 & 590\\
3.510 & 0.00259 &     &      & 240 & 249 &     &    \\
3.540 & 0.00240 & 388 &  400 & 259 & 432 &     &    \\
3.570 & 0.00212 & 422 & 1000 & 281 & 385 & 211 & 610\\
3.585 & 0.00192 &     &      &     &     & 219 & 460\\
3.630 & 0.00170 & 489 &  800 & 326 & 399 &     &    \\
3.690 & 0.00150 & 547 &  500 & 365 & 450 &     &    \\
3.760 & 0.00130 & 637 & 1000 & 424 & 320 & 318 & 477\\
3.820$^*$ & 0.00110 & 710 &  700 &     &     &     &    \\
3.920$^*$ & 0.00092 & 799 & 3280 & 532 & 497 &     &    \\
\cline{1-6}
        &       & \multicolumn{2}{|c|}{$32^3 \times 4$} &
                  \multicolumn{2}{|c|}{$32^3 \times 6$} & & \\
\cline{1-6}
3.820 & 0.00125 &     &      & 480 & 742 & 361 & 509\\
3.920 & 0.00110 & 820 &  550 & 547 & 834 & 410 & 501\\
4.000 & 0.00092 &     &      & 633 & 890 & 475 & 899\\
4.080 & 0.00081 &     &      & 727 & 580 & 549 & 593\\
4.200 & 0.00068 &     &      &     &     & 652 & 300\\
\hline
\end{tabular}
\end{center}
\caption{Coupling constants $\beta$ and light quark masses $m_l$, temperatures $T$ in
MeV and the number of configurations of the given sizes on which screening masses
were computed. The strange quarks have always been ten times heavier than the light
ones, $m_s = 10 m_l$. At the $\beta$ values marked with a star we analyzed
configurations which were generated in an early stage of the project at quark masses
slightly off the LCP.}
\label{tb.runpara}
\end{table}

\section{Free case}                             \label{se.free}

Lattice results are in general affected by discretization as well as by finite volume
effects.  At fixed temperature these are parametrized through the temporal extent of
the lattice, $aT=1/N_\tau$, and the aspect ratio, $T\mathcal{V}^{1/3}=N_s/N_\tau$,
respectively. Discretization and finite volume effects are readily computed
semi-analytically in the non-interacting theory. While in this case both effects are
expected to be larger than in the interacting theory  \cite{Wissel,me} these
computations give at least hints about the sign and the order of magnitude of the
lattice distortions. The free case is also the infinite temperature limit of the
interacting theory. In this limit the spectral function underlying both, temporal as
well as spatial correlation functions, solely consists of a cut in the complex
frequency plane. The free case thus provides a test example for screening masses
arising from a cut.

In the free case at vanishing quark mass only states of one parity contribute to the
correlation functions. In channels $\cM2$ and $\cM5$ to $\cM7$ these are states with
negative parity while in the remaining channels the contributing states have positive
parity. In the latter case an additional phase factor $-(-1)^z$ has been applied to
$C(z)$, see Eq.(\ref{eq.cor-par}) and Table~\ref{tb.phase_factors}, such that for all
channels the effective mass can be defined as
\beq
aM_{eff}(z) = - \ln\lt[ \frac{C(z+1)}{C(z)} \rt].
\label{eq.eff_mass}
\eeq
With this sign convention the correlation functions of all channels are degenerate at
zero quark mass for all values of $N_\tau$ and aspect ratios $N_s/N_\tau\ge4$.  At
large $z$ the spatial meson correlators start oscillating. These oscillations
originate from a contribution $\sim (-)^z \cdot constant$ (in our phase convention)
which vanishes in the thermodynamic limit $N_s \rightarrow \infty$.

\begin{figure*}[!ht]
\bc
\subfigure[]{\label{fig.free_p4}\includegraphics[scale=0.50]{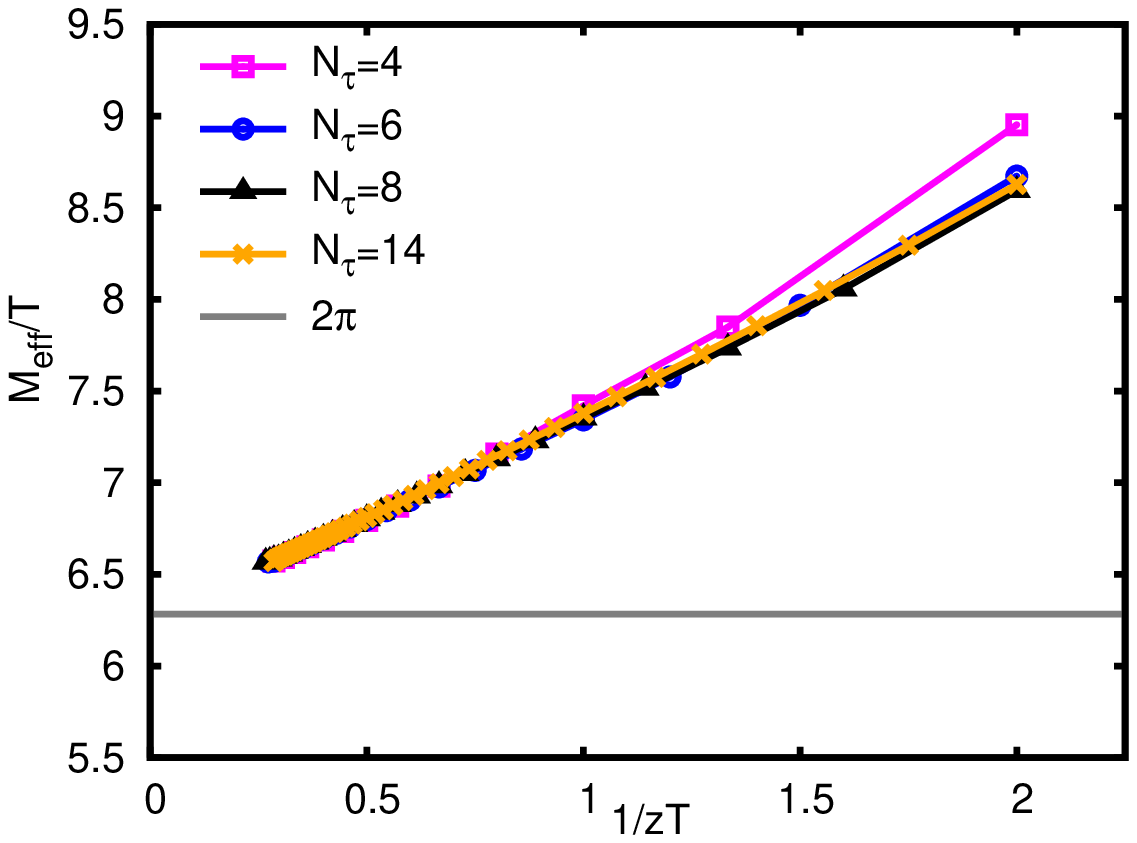}}
\hspace{1.25cm}
\subfigure[]{\label{fig.free_naive}\includegraphics[scale=0.50]{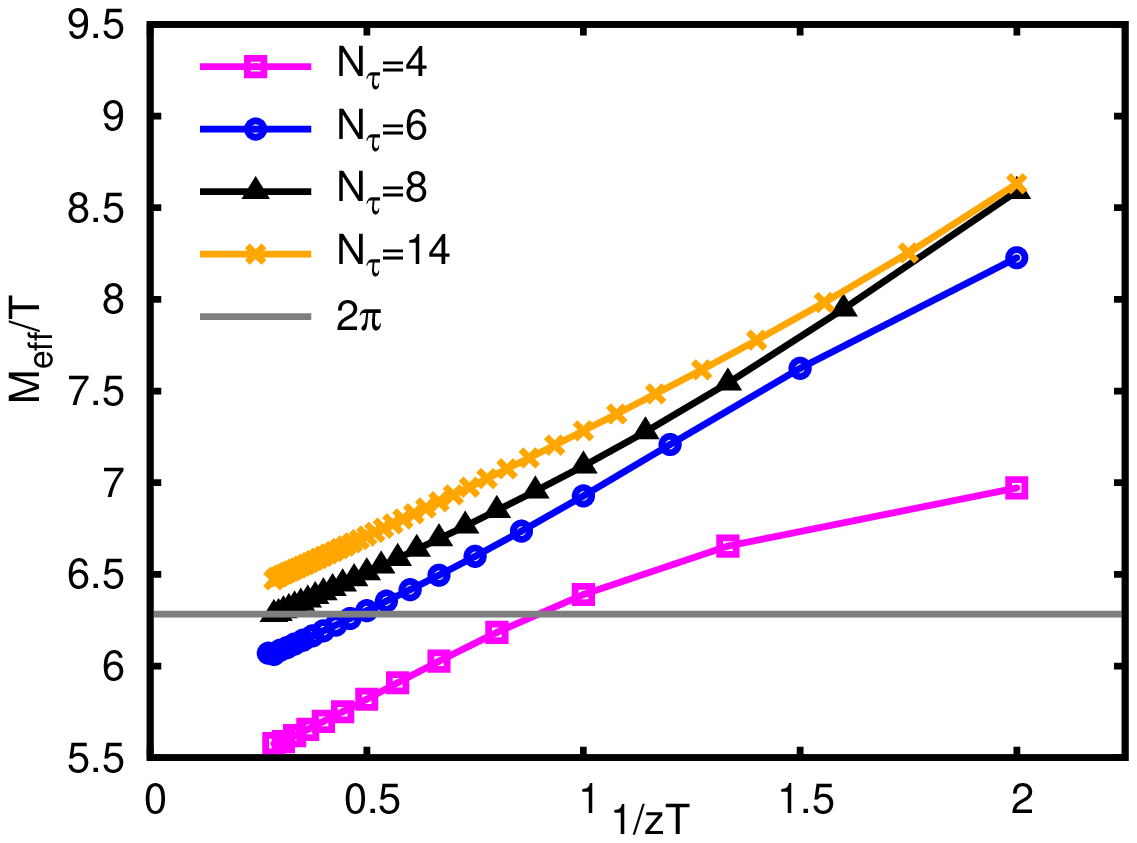}}
\ec
\vspace{-0.3cm}
\caption{Dependence of the effective screening mass, Eq.(\ref{eq.eff_mass}), on
$N_\tau$ and on the separation $zT$, obtained from point correlators for a free
theory with the p4 staggered action (a) and with standard staggered fermions (b).}
\label{fig.free}
\end{figure*}

In Fig. \ref{fig.free_p4} we show $M_{eff}(z)$ as a function of the inverse separation
in units of temperature, $1/(zT)$, for p4 fermions at various values of the temporal
lattice extent. As can be seen, for the improved p4 action there is practically no
$N_\tau$ dependence for $zT \geq 1$.  On the other hand, for the unimproved, standard
staggered action the discretization errors are significantly larger, as is shown in
Fig. \ref{fig.free_naive} for comparison.

However, for both the p4 and the standard staggered action the effective mass from a
point correlator is strongly dependent on $z$, and the value $2\pi T$ is obtained
only in the limit of infinite $z$. This is in accord with the continuum, non-lattice
calculation of the spatial pion correlation function \cite{Friman} where this value
is also only reached in the limit $z \rightarrow \infty$ of the continuum equivalent
of Eq.(\ref{eq.eff_mass}). The $z$ dependence of the effective screening mass is due
to the cut in the spectral function. Because of contributions from quarks at all
momenta, $p_x,p_y$ and all Matsubara frequencies, $\omega_n = \pi T (2n+1)$ which die
out only in the limit $z \rightarrow \infty$ the effective screening mass does not
develop a plateau. This is strictly true for effective masses calculated from
point-to-point correlation functions. Extended sources, for instance wall sources,
suppress contributions from higher quark momenta and thus may lead to a plateau like
behavior in effective mass plots, at least in a limited $z$ interval.  Since infinite
separations can only be achieved in the thermodynamic limit, Fig. \ref{fig.free}
suggests that the screening masses determined from lattice simulations may have
significant dependence on the maximum available distance along which the correlation
functions were measured. This can be considered as a finite volume effect, although
an indirect one as for aspect ratios larger than 3 there is practically no volume
dependence of the free results other than through the limited $z$ range that is
available. As a corollary, we note that the absence of a plateau in effective masses
extracted from point correlators is a necessary condition to the verification of free
field behavior in screening masses.

\section{Results}                          \label{se.light}

The following results on screening masses of mesonic states built from light quarks
have been obtained on dynamical $N_F=2 + 1$ gauge field configurations.  The analysis
was performed on lattices of size $16^3 \times 4$, $24^3 \times 6$ and $32^3 \times
8$, \cite{spatial-string,RBCBi-eos,hotQCD} accompanied by some cross-checks on
available high temperature lattices with larger aspect ratios at $N_\tau = 4$ and
$6$, (see Table \ref{tb.runpara}).  On all those configurations we computed spatial
correlations of meson operators (see Table~\ref{tb.phase_factors}) made out of
degenerate u and d quarks. In addition we studied mesons with $\bar u s$ and $\bar s
s$ flavor content on the lattices with temporal extent $N_\tau = 4$ and $6$.  In all
cases the quark masses of the quark propagators have been chosen to be the same as in
the quark determinant, i.e. in the generation of the configurations.  On all lattices
we computed point-to-point correlation functions.  Subsequently, for most of the
$N_\tau = 4$ and 6 configurations, and some of the ones at $N_\tau = 8$, (see the
tables in the appendix) we also analyzed correlators from wall sources.

In general, in a staggered fermion meson (screening) correlator a meson is
accompanied by an opposite parity meson, see Eq. (\ref{eq.cor-par}). However, in the
present study, in most channels we have seen indications for the presence of both of
these mesons at most at the lowest temperatures. With increasing temperature, the
contribution of one of the two mesons died out very fast.  Hence, in most cases we
found the signature of only one of the states. Specifically:

\noindent -- The $\cM1$ channel is dominated by the oscillating positive parity
scalar.  A non-oscillating negative parity contribution due to a pseudo-scalar state
could not be isolated.  We will denote the screening masses in this scalar channel as
$M_{SC}$, keeping in mind that at least at zero temperature the scalar channel
contains two pseudo-scalar meson contributions which may become particularly
important in the presence of taste violations \cite{MILCtaste}.  

\noindent -- In the $\cM2$ channel we have not seen any oscillating contribution.
This channel is dominated by the Goldstone pseudo-scalar at all temperatures.  The
screening masses in this channel will be denoted by $M_{PS}$.  

\noindent -- The $\cM3$ and $\cM4$ point correlators are noisy at low temperature but
the signal-to-noise ratio improves greatly with rising $T$.  Wall source correlators
delivered suitable signals at all temperatures.  In all fits we have found stable
signatures only of positive parity axial-vectors.  The screening masses of the
axial-vectors, denoted by $M_{AV}$ later, coming from these two channels are
degenerate

\noindent -- In $\cM6$ and $\cM7$ we found only the negative parity vector states.
The signal-to-noise ratios were similar to the ones seen in $\cM3$ and $\cM4$. The
screening masses of these states, $M_V$, coming from these two channels are found to
be degenerate for all temperatures.  

\noindent -- Channel $\cM5$ as well as $\cM8$ were found to be very noisy in point
correlators and reasonable signals could only be obtained from them at the highest
temperatures. The wall correlators are at low temperature dominated by the
axial-vector in $\cM5$ channel and the vector in $\cM8$ channel. At high temperature,
independent of the source we found signatures of vector as well as axial-vector
states in both channels. 

\noindent -- These observations hold for all flavor combinations where, however, the
signal-to-noise ratio is in general slightly better for the heavier combinations.

\noindent -- Contrary to the free case, in none of the correlation functions a
contribution constant in $z$ was observed.

\begin{figure*}[!ht]
\subfigure[]{\label{fig.ps}\includegraphics[scale=0.50]{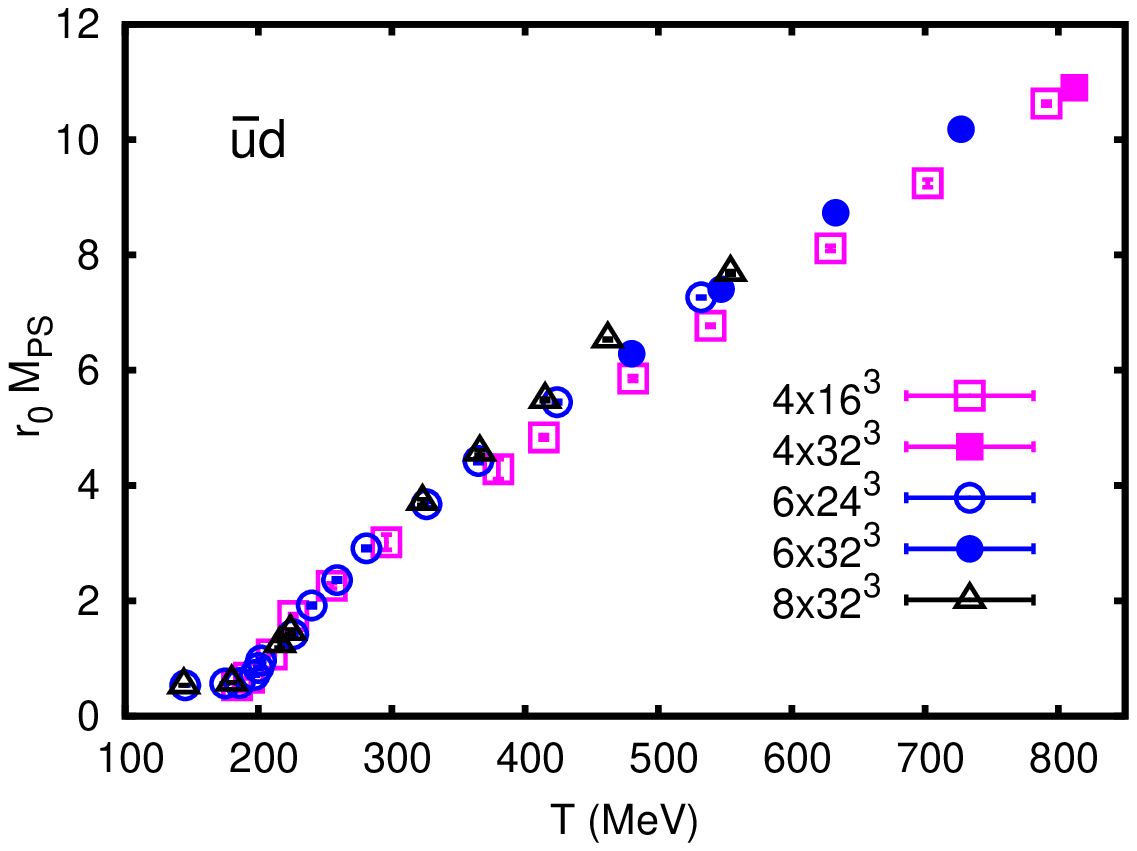}}
\subfigure[]{\label{fig.psflav}\includegraphics[scale=0.50]{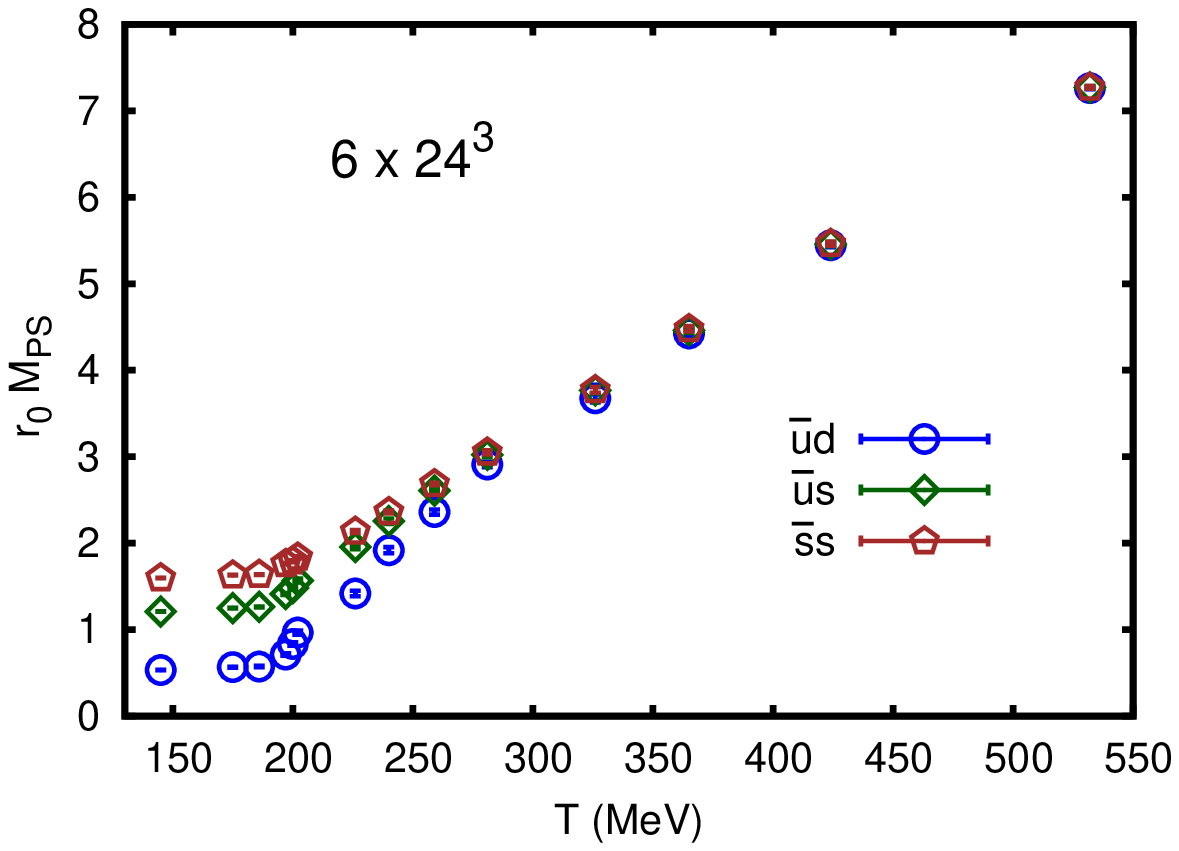}}
\subfigure[]{\label{fig.t0_comp}\includegraphics[scale=0.50]{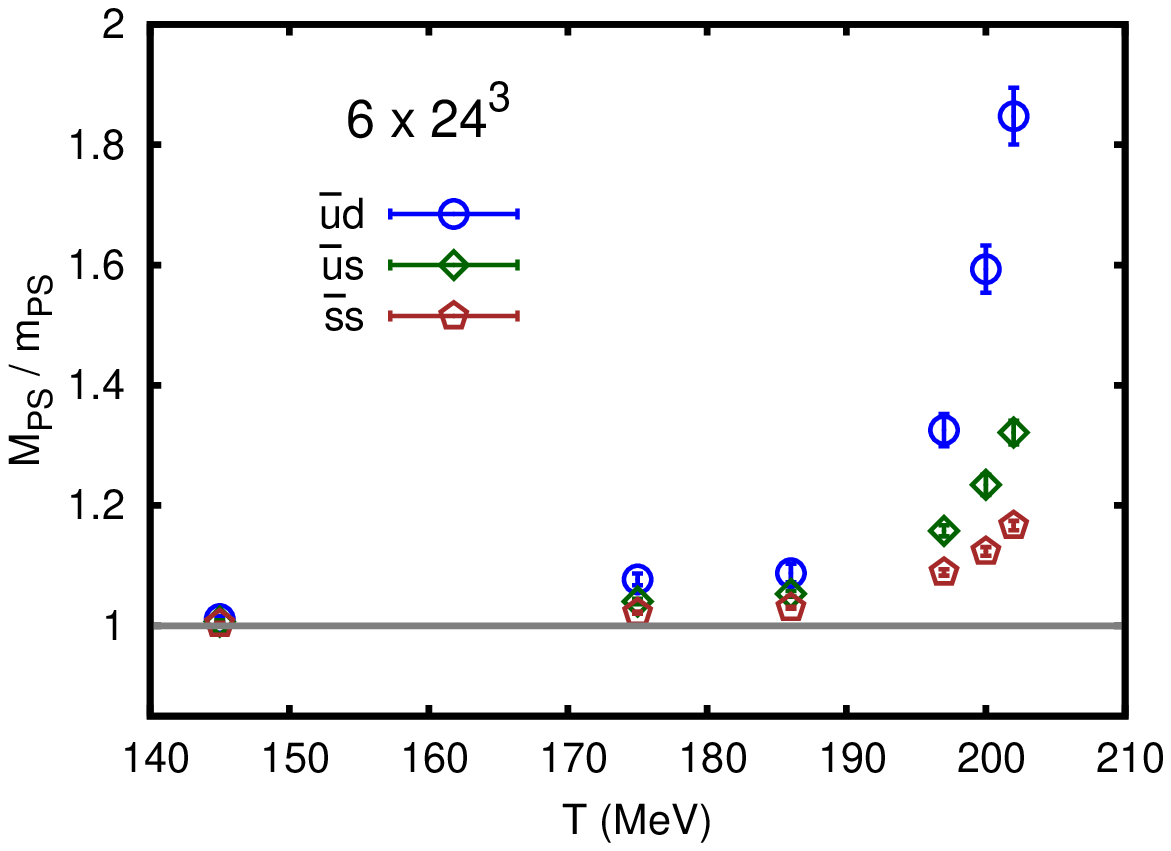}}
\caption{(a) The pseudo-scalar screening masses in units of $r_0$ for $N_\tau = 4,6$
and 8. The $N_\tau=4$ temperatures have been shifted by -8 MeV and the $N_\tau=8$
ones by +5 MeV, see text.  (b) Flavor dependence of the pseudo-scalar screening masses
at $N_\tau = 6$. (c) Ratio of the pseudo-scalar screening masses at $N_\tau=6$ to the
corresponding zero temperature masses determined at the same couplings along the line
of constant physics with $m_\pi\simeq 220$ MeV and $m_K\simeq500$ MeV.}
\end{figure*}

\begin{figure*}[!ht]
\subfigure[]{\label{fig.sc}\includegraphics[scale=0.50]{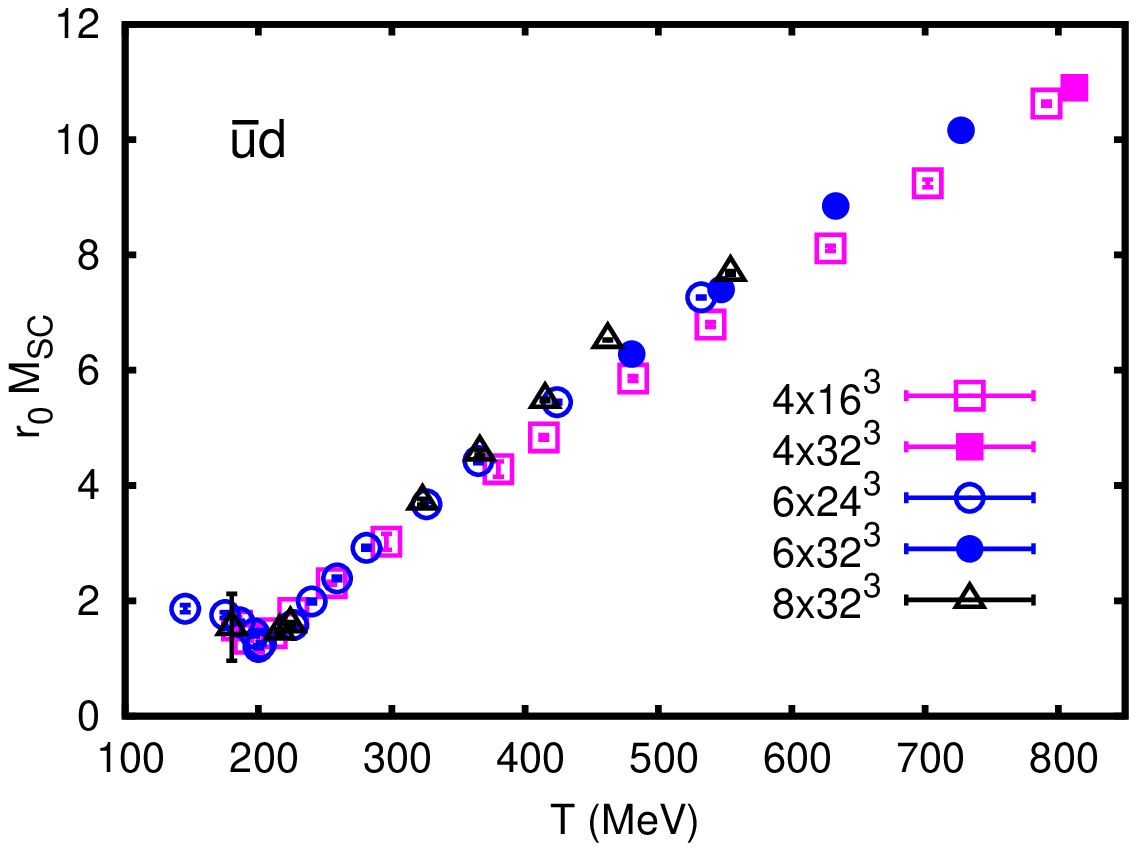}}
\subfigure[]{\label{fig.scflav}\includegraphics[scale=0.50]{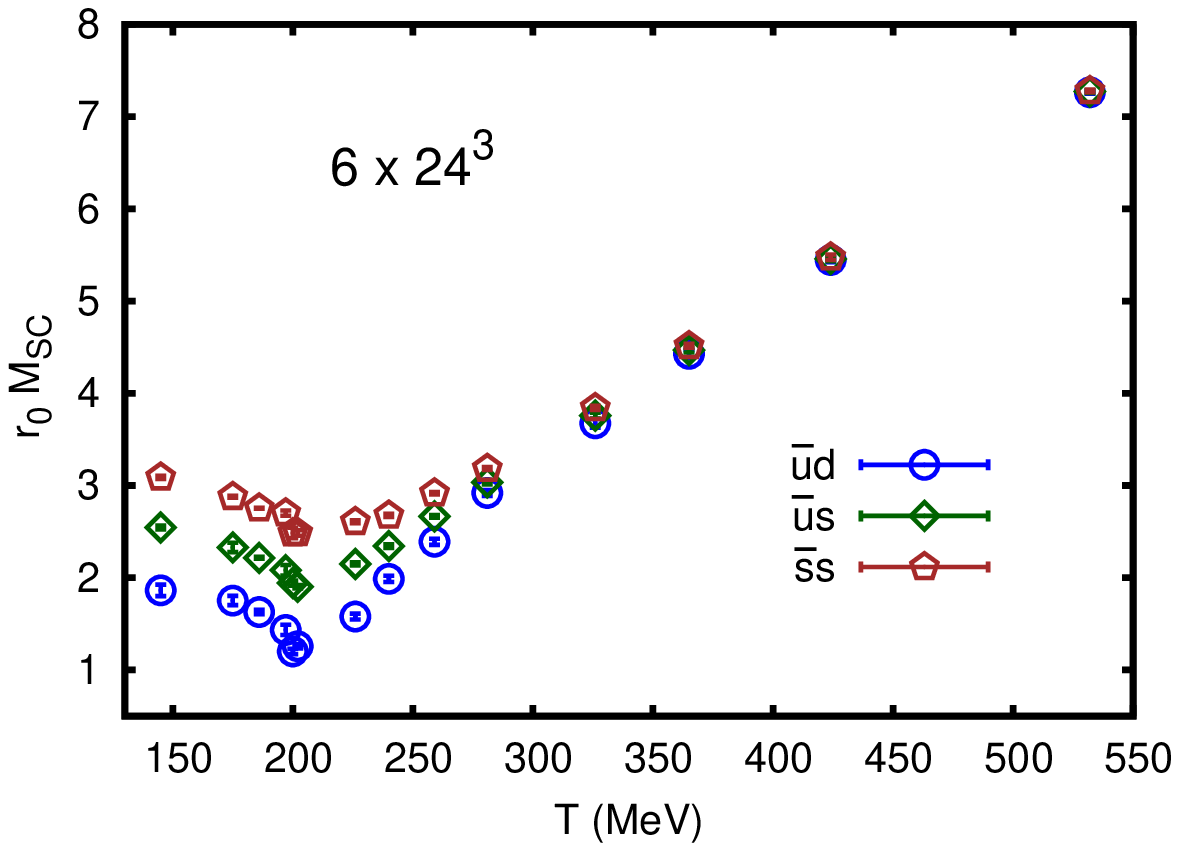}}
\subfigure[]{\label{fig.ps-sc_comp}\includegraphics[scale=0.50]{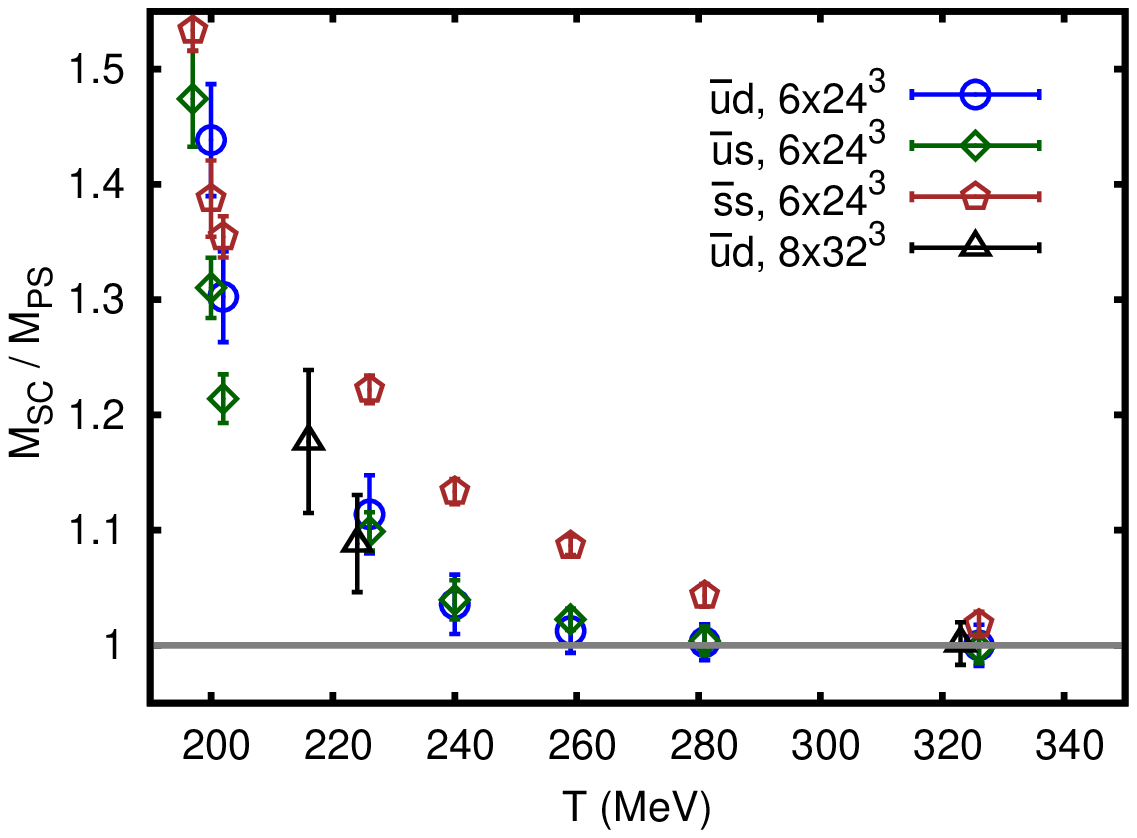}}
\caption{(a) The scalar screening masses in units of $r_0$ for $N_\tau = 4,6$ and 8.
Temperatures have been shifted as in Fig. \ref{fig.ps}.  (b) Flavor dependence of the
scalar screening masses at $N_\tau = 6$.  (c) Ratio of the scalar to the
pseudo-scalar screening mass at high temperature. The data is from the $N_\tau = 6$
and $8$ lattices for the $\bar u d$ flavor combination and from $N_\tau=6$ for $\bar
u s$ and $\bar s s$.}
\end{figure*}

In all cases where acceptable signal-to-noise ratios were obtained within our
statistics, the effective masses $M_{eff}(z)$ from point-to-point correlators show
reasonable plateaux in $z$ within errors. This is particularly true at high
temperature where we also analyzed lattices with aspect ratios larger than 4 to
access larger $zT$ values.  There is a slight trend in the effective masses to
decrease with increasing $z$, but in none of the channels the free field behavior as
depicted in Fig. \ref{fig.free_p4} could be verified.

Subsequently we analyzed spatial meson correlators from wall sources built along the
$x,y,\tau$ directions.  As pointed out in the previous section, effective masses from
wall source correlators potentially pretend the presence of genuine bound states even
in the free case because contributions from free quarks with higher momenta are
suppressed by construction.  However, the point-to-point correlation functions have
already indicated the absence of a substantial contribution from a free quark cut at
the temperatures investigated.

The effective masses from the wall source correlators in general approach plateaux
from below, as at zero temperature.  For the scalar and the pseudo-scalar channels the
error bars on the screening masses are of the same magnitude as for the point
correlators. The wall source masses tend to be slightly lower than the point source
ones but the differences are at the one $\sigma$ margin. Since the scalar and
pseudo-scalar data set for point sources is most complete we will plot point source
results in the following, for wall source numbers we refer to the tables in the
appendix.  In the vector and axial-vector channels the signal-to-noise ratio is much
better for the wall sources, in particular at low temperatures.  At high temperatures
the differences between wall and point source results are again small, with somewhat
smaller error bars on the wall source results.  Except for the $N_\tau=8$ lattices,
we will therefore plot and discuss wall source results for those channels below.

Independent of the source, the screening mass values which will be quoted in the
following have been obtained from fits to the correlators with a minimum distance
$z_{min}$ from source and sink of about $z_{min} T \simeq 1$, except for the high
temperature checks on lattices with aspect ratios larger than 4 where larger
separations up to $z_{min}T \simeq 2$ could be accessed. Errors were determined by
jack-knifing the fits.

For the purpose of showing our data as a function of a common temperature scale,
throughout this paper the data for $N_\tau = 4$ has been shifted in temperature by a
value of $\Delta T = - 8$ MeV and the one for $N_\tau = 8$ by $\Delta T = + 5$ MeV to
properly account for the $N_\tau$ dependence of the transition region \cite{hotQCD}.
Also throughout this paper we will use the chiral crossover temperature of
$T_c\simeq196$ MeV as our reference transition temperature. This corresponds to the
chiral crossover temperature, determined from the peak of the disconnected chiral
susceptibility, for $6\times24^3$ lattices with $m_\pi\simeq220$ MeV and
$m_K\simeq500$ MeV \cite{hotQCD}.

In Fig. \ref{fig.ps} we show our results for $M_{PS}$, in the $\bar u d$ flavor
combination, in units of $r_0$ for all $N_\tau$ values of 4, 6 and 8 as a function of
temperature. The pseudo-scalar screening masses stay constant to a high degree of
accuracy below the transition region and rise rapidly above.  For instance, at a
temperature of about 250 MeV the (lightest) pseudo-scalar mass has already reached a
value of $\simeq 3.5 T$ and at $T \simeq 400$ MeV the value has increased to $\simeq
5.25 T$ in temperature units. The rapid rise of $M_{PS}/T$ indicates that in this
temperature range the scale for the pseudo-scalar screening mass is not set solely by
the temperature.

\begin{figure*}[!ht]
\subfigure[]{\label{fig.v}\includegraphics[scale=0.50]{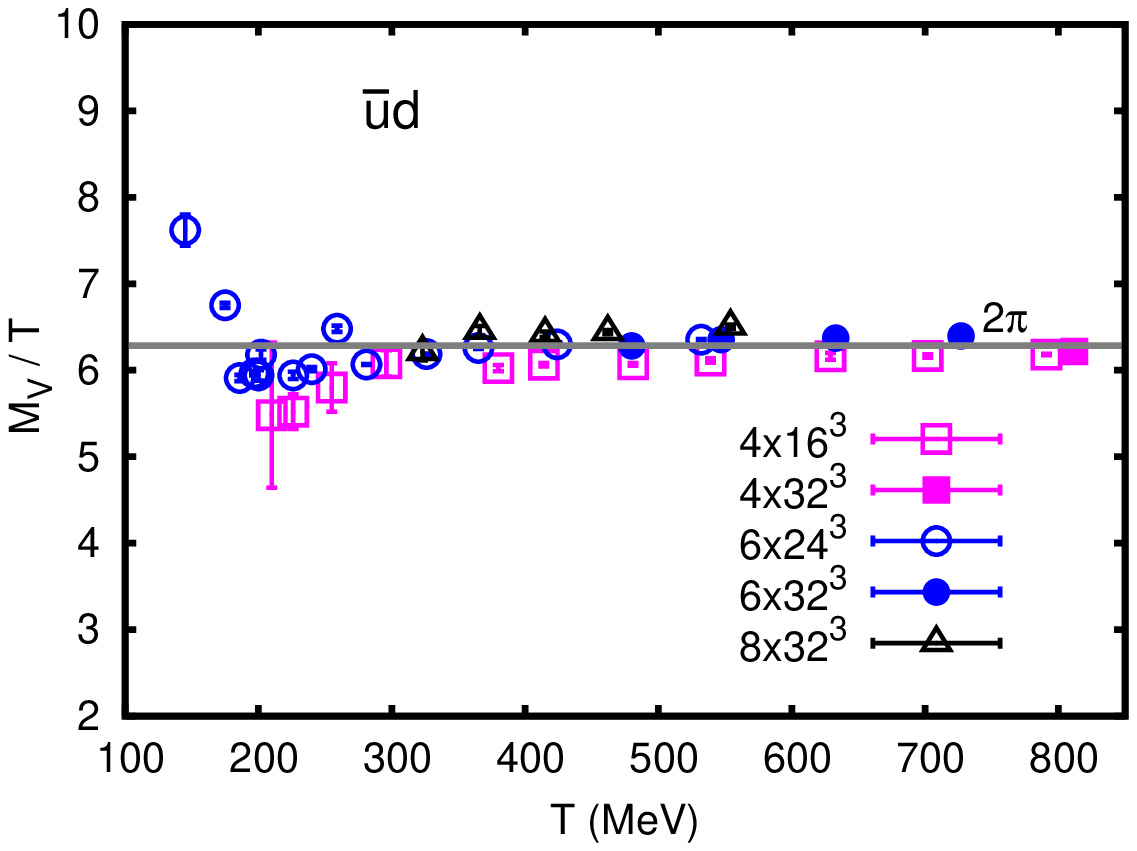}} 
\subfigure[]{\label{fig.vflav}\includegraphics[scale=0.50]{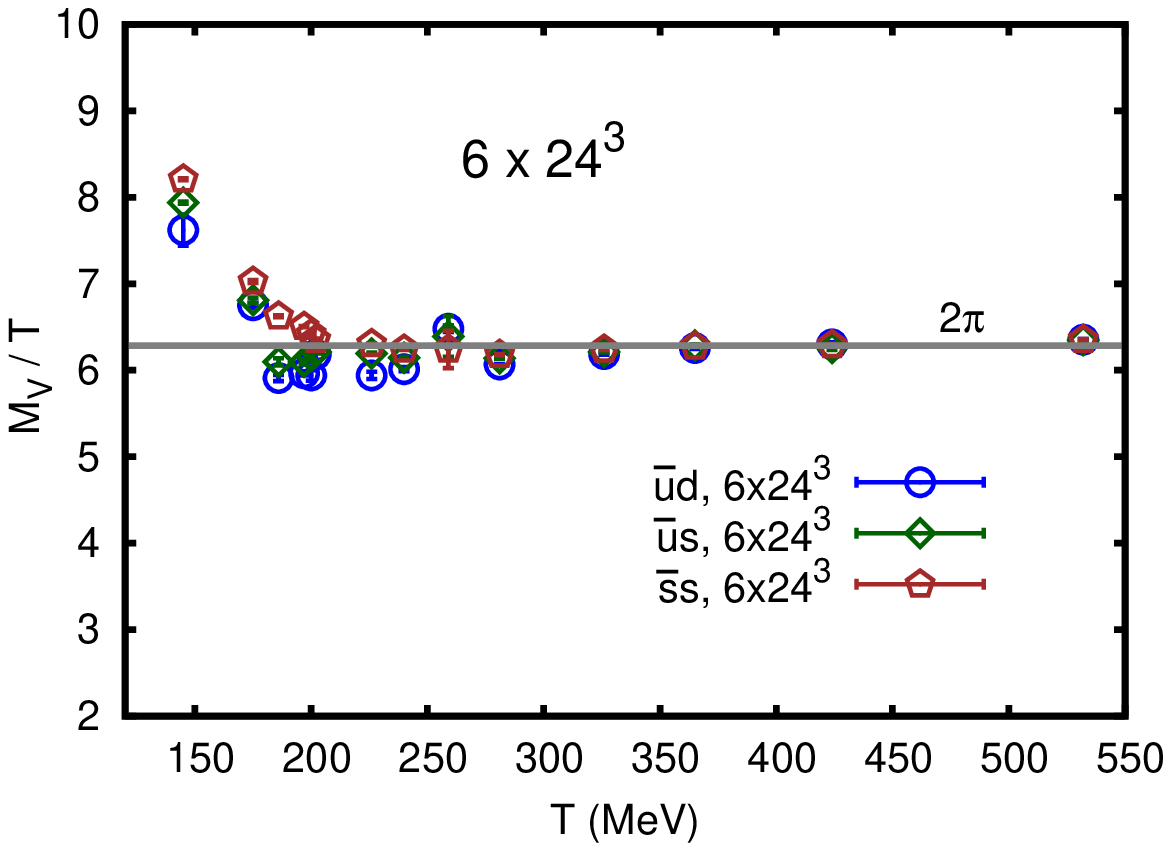}}
\subfigure[]{\label{fig.ps_v}\includegraphics[scale=0.50]{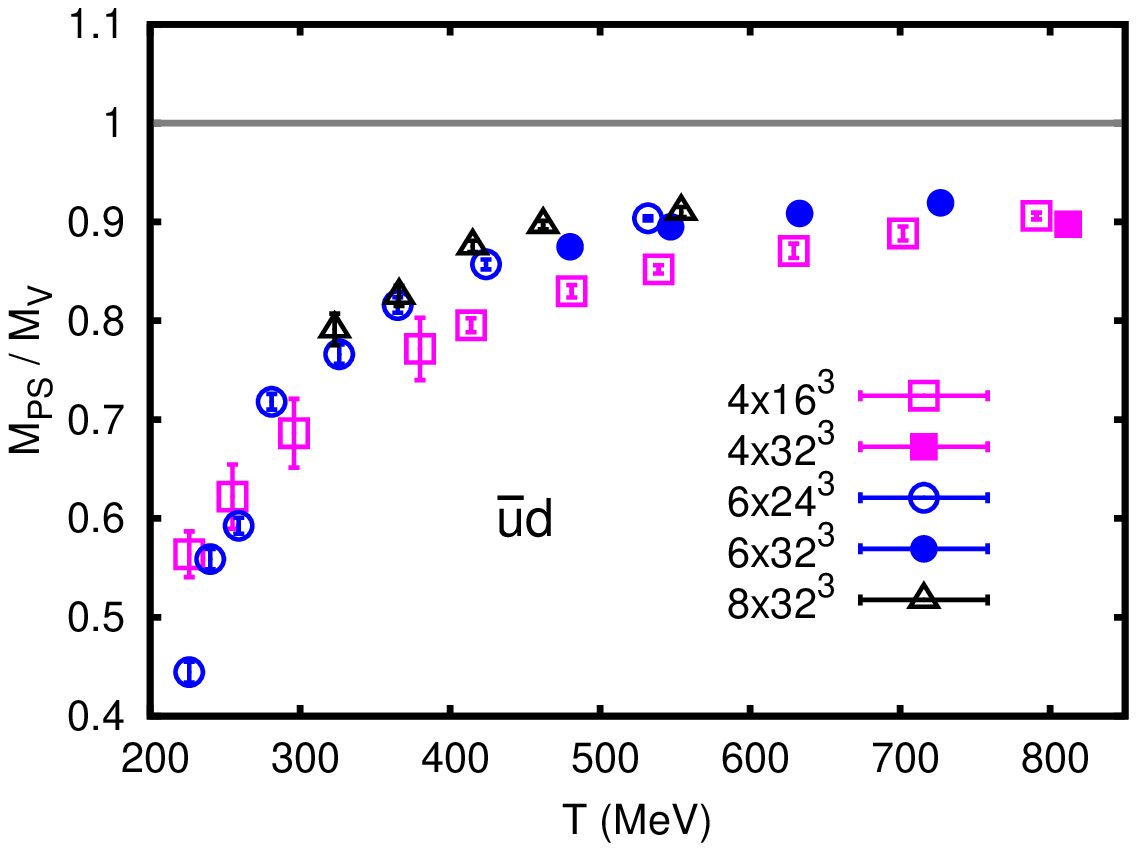}}
\caption{(a) Temperature and lattice spacing dependence of the vector screening mass,
$M_{V}/T$, in units of temperature for the $\bar u d$ flavor combination.
Temperatures have been shifted as in Fig. \ref{fig.ps}. (b) Flavor dependence of the
vector screening masses at $N_\tau = 6$.  (c) Ratio of the pseudo-scalar to vector
screening mass in the $\bar u d$ flavor combination.}
\label{fig.vtotal}
\end{figure*}

In Fig. \ref{fig.psflav} we compare the different flavor channels for $N_\tau = 6$.
Below the transition region, for all flavor combinations the pseudo-scalar screening
masses stay approximately constant in temperature and are markedly different from
each other, in accord with the mass differences at zero temperature.  Above the
transition region the differences between the flavor channels become smaller with
rising temperature and practically vanish at around 300~MeV.

The differences between zero temperature masses and finite temperature screening
masses in the confined phase are inspected more closely in Fig. \ref{fig.t0_comp}.  As
can be seen, the ratio of $M_{PS}$ to $m_{PS}$, the ordinary zero temperature
Goldstone pion masses determined at the same couplings along the line of constant
physics, starts differing from one for temperatures $T\gtrsim170$ MeV. This holds true
for all the three flavor combinations $\bar ud$, $\bar us$ and $\bar ss$.  However,
in the range $170~{\rm MeV}\le T \lesssim 190$ MeV, the differences are distinct but
only of the order of $5-10\%$ slightly depending on the quark mass. 

In Fig. \ref{fig.sc} we show the temperature dependence of $M_{SC}$, normalized by
$r_0$, in the $\bar u d$ flavor combination for our lattices with temporal extents 4,
6 and 8. Note that $M_{SC}$ decreases with $T$ at low temperature before it starts to
rise rapidly at $200$ MeV.  This is seen more clearly in Fig. \ref{fig.scflav} where
we plot $M_{SC}$ from the $N_\tau = 6$ lattices for all flavor combinations.  Above
$200$ MeV the screening mass differences between the flavor channels persist up to
about 325 MeV and vanish at higher temperature.

In the presence of sufficiently strong taste violations and at small enough Goldstone
boson masses, at least at zero temperature the scalar flavor non-singlet channel,
$a_0$, may receive its dominant contribution from an unphysical two-Goldstone boson
threshold \cite{MILCtaste}. However, in our case of screening correlators we observe
that at temperatures in the confined phase $M_{SC}$ as obtained from fits to the data
is significantly larger than two times the Goldstone boson screening mass.  At the
same time $M_{SC}$ is lower than twice the Kaon mass as well as the $\pi \eta$
threshold where the $\eta$ mass was estimated with the Gell-Mann--Okubo formula
$m_\eta^2 = (4m_K^2 - m_\pi^2)/3$.  This indicates that here indeed $M_{SC}$ denotes
the screening mass of a genuine scalar.  We then compare this scalar screening mass
with the pseudo-scalar one at temperatures above $200$ MeV in Fig.
\ref{fig.ps-sc_comp}.  The figure highlights that the differences between the two
gradually disappear with rising temperature, leading to degeneracy, but only above a
temperature of 250 MeV.  Note that the differences for the channels containing light
quarks, i.e. the one with the light $\bar u d$ content as well as the heavier $\bar u
s$ combination, agree with each other to a good degree of accuracy suggesting that
those differences survive in the light quark chiral limit. The non-degeneracy of
scalar and pseudo-scalar screening mass still observed at temperatures up to at least
250 MeV thus indicates that the anomalous $U_A(1)$ symmetry is not effectively
restored at this temperature. The non-zero mass splitting $(M_{SC}-M_{PS})$ for the
$\bar ss$ sector for $250\ \mathrm{MeV}\lesssim T\lesssim325\ \mathrm{MeV}$ is most
likely due to larger explicit $U_A(1)$ breaking in presence of the larger strange
quark masses.  This is also supported by the fact that also in this sector this mass
splitting vanishes for $T\gtrsim325$ MeV which, as mentioned before, is the
temperature range where the difference between the light and strange quark masses
become unimportant.

\begin{figure*}[!ht]
\subfigure[]{\label{fig.av}\includegraphics[scale=0.50]{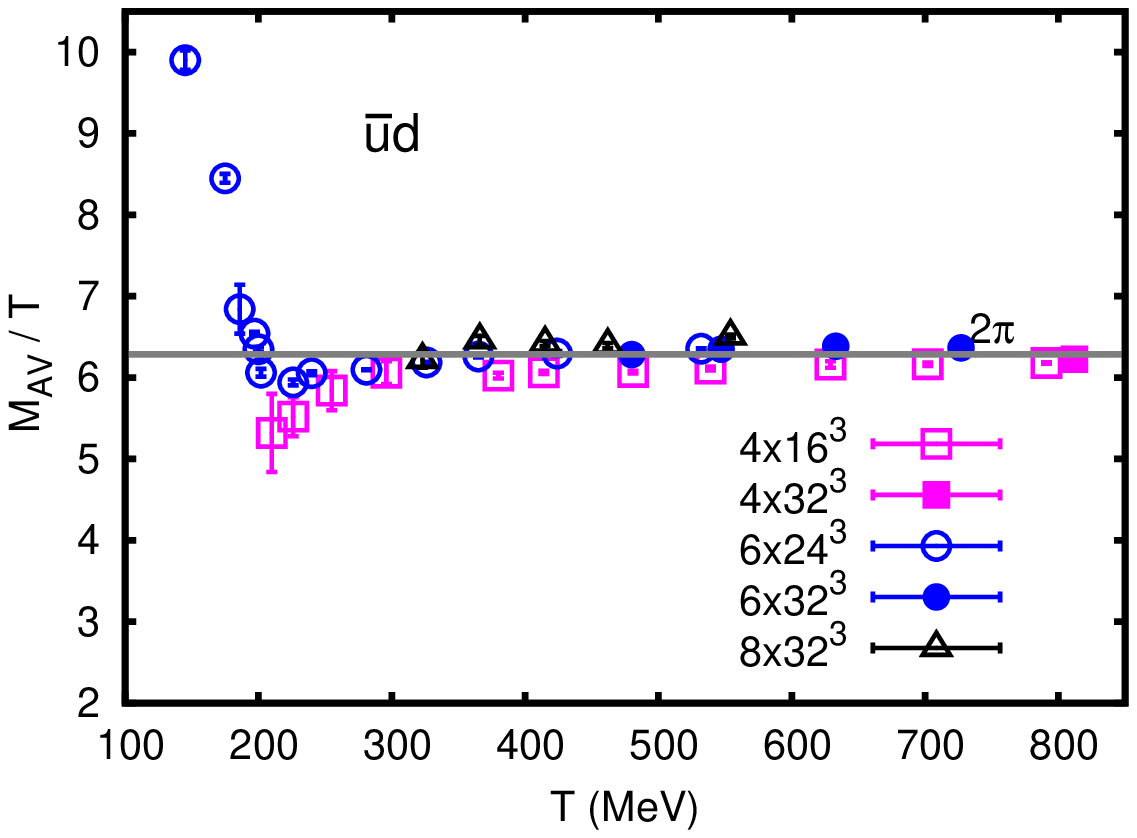}}
\subfigure[]{\label{fig.avflav}\includegraphics[scale=0.50]{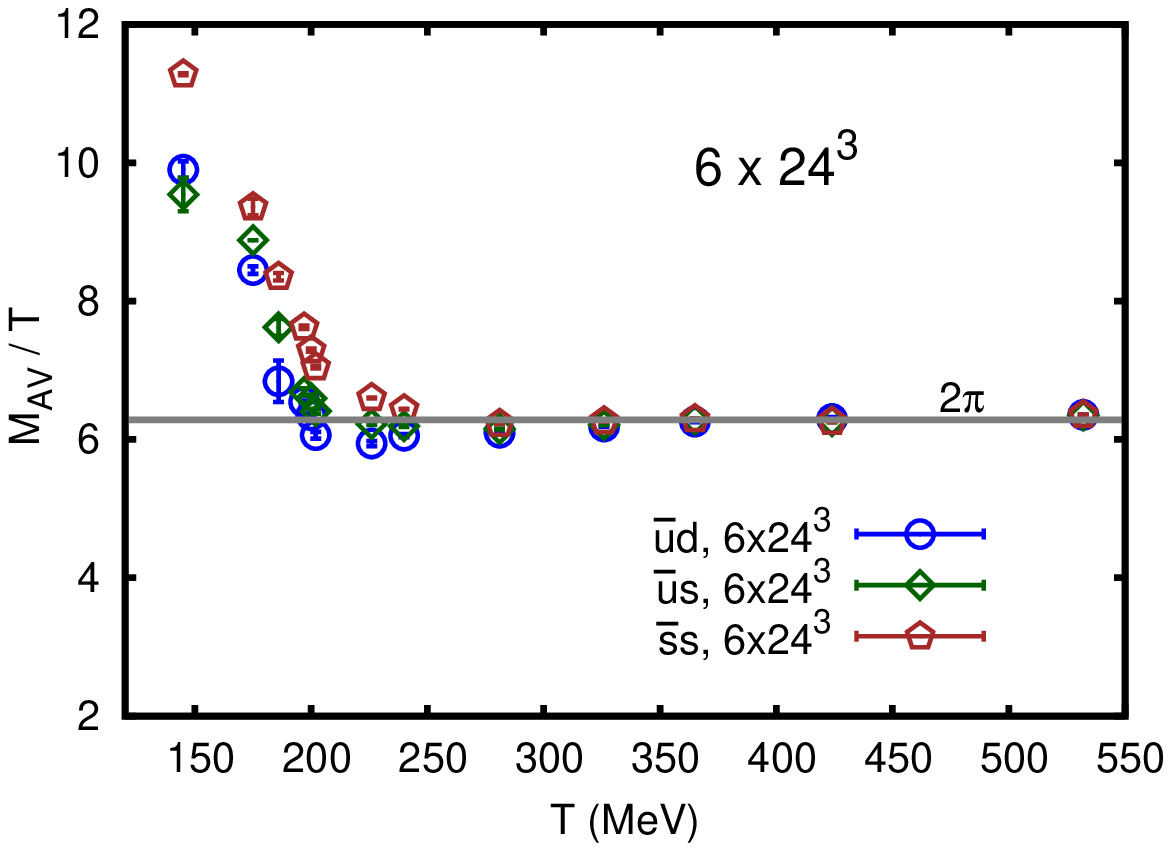}}
\subfigure[]{\label{fig.avrat}\includegraphics[scale=0.50]{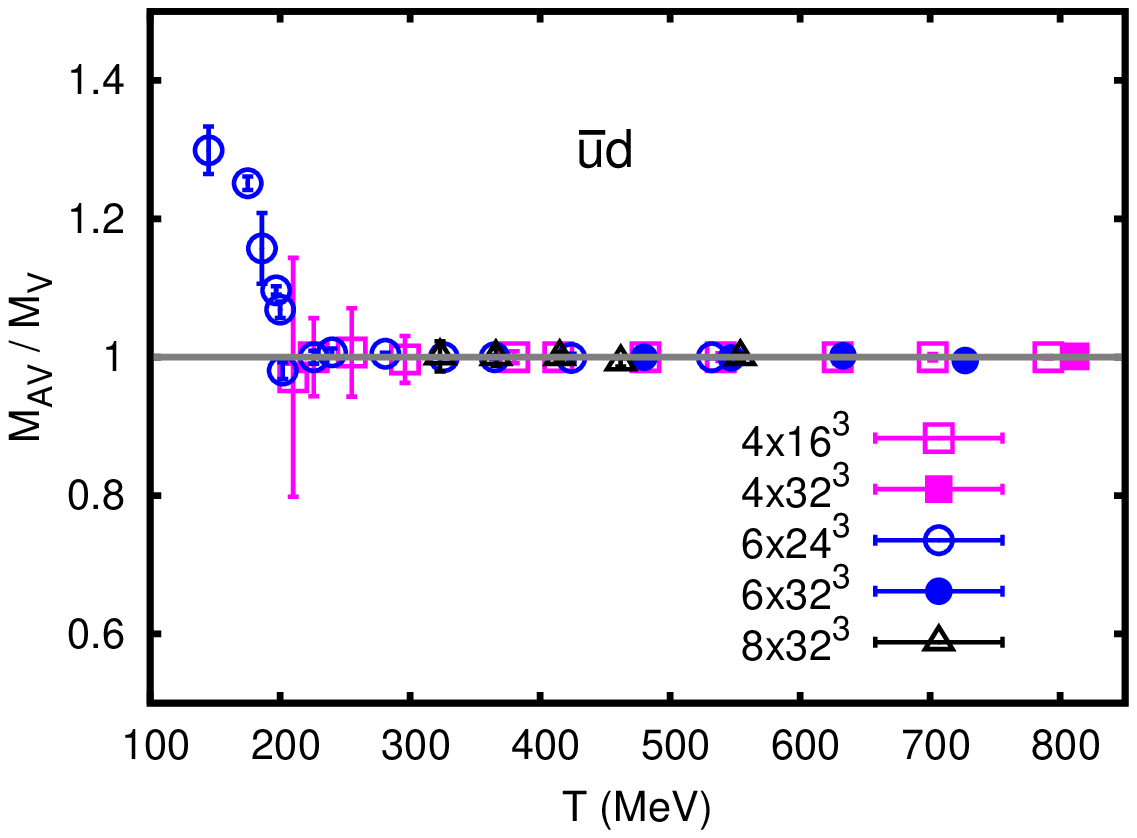}}
\caption{The axial-vector screening mass: (a) Temperature and lattice spacing
dependence in the $\bar u d$ flavor combination, as in Fig. \ref{fig.v}.  (b) Flavor
dependence at $N_\tau=6$. (c) Ratio to the vector screening mass, in the $\bar u d$
flavor combination.}
\label{fig.avtotal}
\end{figure*}

In Fig. \ref{fig.v} the results for $M_V$ at different $N_\tau$ values are shown.  At
temperatures below the transition the vector screening mass $M_V$ in physical units
is roughly constant in $T$, while for temperatures larger than about $250$ MeV its
ratio with $T$, $M_V/T$, is approximately constant over the investigated temperature
range. This indicates that different from the pseudo-scalar and scalar channels the
temperature sets the scale for the vector screening masses already slightly above the
transition. The ratio $M_V/T$ is slightly larger than the continuum free result at
infinite separation. However, a comparison with the free lattice result in the region
around $zT \simeq 1$ which typically is the lower boundary of the fit window shows
that the data are about $20\%$ lower than this number. Moreover, even at the highest
temperatures available the effective masses from point correlators do not show the
$z$ dependence as in the free case.

Fig. \ref{fig.vflav} shows the flavor dependence of the vector screening masses as a
function of temperature for $N_\tau = 6$. At low temperatures the screening masses
are ordered according to the strange quark content whereas already at a temperature
of $T \simeq 225$ MeV, the vector screening masses are flavor independent, supporting
that for the vector channel it is the temperature which sets the scale.

We also inspect the degeneracy of $M_{PS}$ and $M_{V}$, which is expected at least in
the limit of infinite temperature.  In Fig.\ \ref{fig.ps_v} we show the ratio
$M_{PS}/M_{V}$ for the lightest flavor combination, noting that the ratio is
independent of the quark flavor for temperatures larger than 300 MeV.  Also for this
ratio, the results from the $N_\tau=4$ lattices are somewhat lower than those from
the other two lattice sets with  $N_\tau = 6$ and 8, which are in good agreement with
each other.  Overall we find that $M_{PS}/M_V$ is statistically significantly lower
than $1$ even at temperatures $T\gtrsim500$ MeV. Taking the $N_\tau=8$ result at
$T=550$~MeV at face value the difference from unity is about $9\%$. There is,
however, a slight trend that the ratio rises with temperature. 

Fig. \ref{fig.avtotal} summarizes our results for the axial-vector screening masses.
Below the transition the screening masses are ordered according to their strangeness
content but above the flavor dependence is weak, Fig. \ref{fig.avflav}. Note that
below $200$ MeV the axial-vector screening mass is decreasing with rising temperature
leading to a decrease in its ratio to the vector screening mass, Fig.
\ref{fig.avrat}. This ratio becomes compatible with $1$ around $T\approx200$ MeV.
Thus within errors axial-vector and vector screening masses become degenerate at this
temperature and remain so at all larger temperatures.  Although the quark masses are
not at the chiral limit the restoration of chiral symmetry is thus also reflected in
spectrum results. 

The longitudinal $\cM5$ and $\cM8$ channels in general have the largest error bars,
also when wall sources are used. In the wall source correlators at temperatures below
the transition, $\cM5$ is dominated by an axial-vector while in $\cM8$ only a vector
could be isolated. Both states are, within errors, degenerate with the transverse
ones. At high temperature, in wall as well as point correlators, where available, the
vector contribution to $\cM5$ has the same screening mass as the axial-vector
contribution to $\cM8$, $M_-(\cM5)=M_{+}(\cM8)$. Both values are about $5\%$ larger
than the degenerate transverse screening masses, $M_{AV}$ or $M_V$.  Similarly, the
axial-vector contribution to $\cM5$ is equal to the vector contribution to $\cM8$,
$M_{+}(\cM5)=M_-(\cM8)$. These screening masses turn out to be roughly $20\%$ larger
than $M_V$. As already noted in Section~\ref{se.operators}, due to the breakdown of
$O(3)$ rotational symmetry to $O(2)\times Z(2)$ for spatial correlators at finite
temperature \cite{gupta} the longitudinally polarized (axial-) vector screening
states in $\cM5$ and $\cM8$ do not need to be degenerate with the transverse
polarizations of the V and AV channels. However, an explanation for the ordering
observed here for the first time is not known to us.

\section{Summary and discussion}               \label{se.summary}

Utilizing the gauge configurations generated by the RBC-Bielefeld
\cite{spatial-string,RBCBi-eos} and the HotQCD \cite{hotQCD} collaborations using the
improved p4 fermion action along a line of constant physics, determined by
$m_\pi\simeq 220$ MeV and $m_K\simeq 500$ MeV, we have investigated the screening
masses of mesons in $2+1$ flavor QCD from $8$ different local meson operators listed
in Table\ \ref{tb.phase_factors}. In order to address discretization errors the
analysis was carried out at three different values of the temporal lattice extent,
$N_\tau=4,\ 6$ and $8$. Although for most of our analyses we have used lattices with
volumes $T\mathcal{V}^{1/3}=4$, at high temperatures we have also analyzed some
lattices with larger volumes in order to access larger separations. Moreover, the
quark mass dependence was studied by analyzing screening correlators for three
different quark flavor combinations, \viz $\bar ud$, $\bar us$ and $\bar ss$. In
order to gain some idea about the systematics we have computed correlation functions
using both point and wall sources. 

\begin{figure}[!t]
\bc
\includegraphics[scale=0.60]{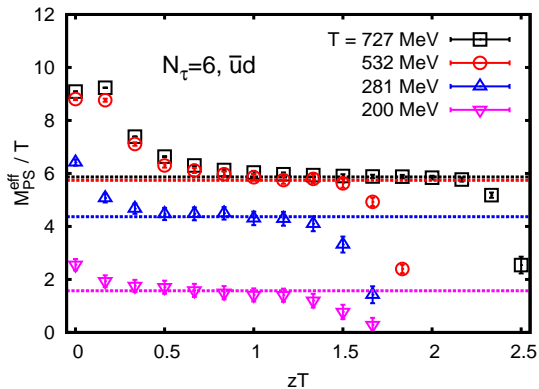}
\ec
\vspace{-0.25cm}
\caption{Effective screening masses in the pseudo-scalar channel as a function of the
distance $zT$ for some illustrative values of temperatures. The dashed lines indicate
the values of the respective screening masses obtained from $\cosh$ fits to the
correlators. Except for the highest temperature, where the correlator was measured
using wall source, all the correlators were obtained using point sources on
$N_\tau=6$ lattices for the $\bar ud$ flavor combination.} 
\label{fig.meff_ps}
\end{figure}

As discussed in Section \ref{se.free}, the presence of a cut in the spectral function
will be manifested through a decreasing effective screening mass (coming from point
source correlation function) as a function of increasing $zT$. Such a
non-plateau-like behavior is a characteristic of the free theory. However, within our
present level of accuracy and an available range of $1<zT<2.5$, we could not identify
a clear decrease of the effective screening masses with increasing $zT$ and have
rather observed plateau-like behaviors. As an illustration, in Fig.\
\ref{fig.meff_ps} we show the effective screening masses in the pseudo-scalar channel
as a function of $zT$ for a large range of temperatures. As can be seen from this
figure, we have not found any clear evidence of a non-plateau-like behavior in the
effective screening masses as seen in the free case (see Fig.\ \ref{fig.free}).
Moreover, the plateau-like structure in the effective screening masses are in good
agreement with the values of the screening masses that have been obtained by fitting
those correlators.

In the low temperature region of $0.75T_c\lesssim T\lesssim T_c$ we have found that
the PS screening mass remains almost constant (within $<10\%$) staying close to the
corresponding zero temperature mass and then starts rising rapidly. In this low $T$
regime also the V screening mass remains almost constant and then shows a rapid
increase. On the other hand, the SC and the AV scalar screening masses show
decreasing tendencies, attain minima around $T_c$ and then start rising again.

\begin{figure*}[!ht]
\subfigure[]{\label{fig.corr-v-av}\includegraphics[scale=0.50]{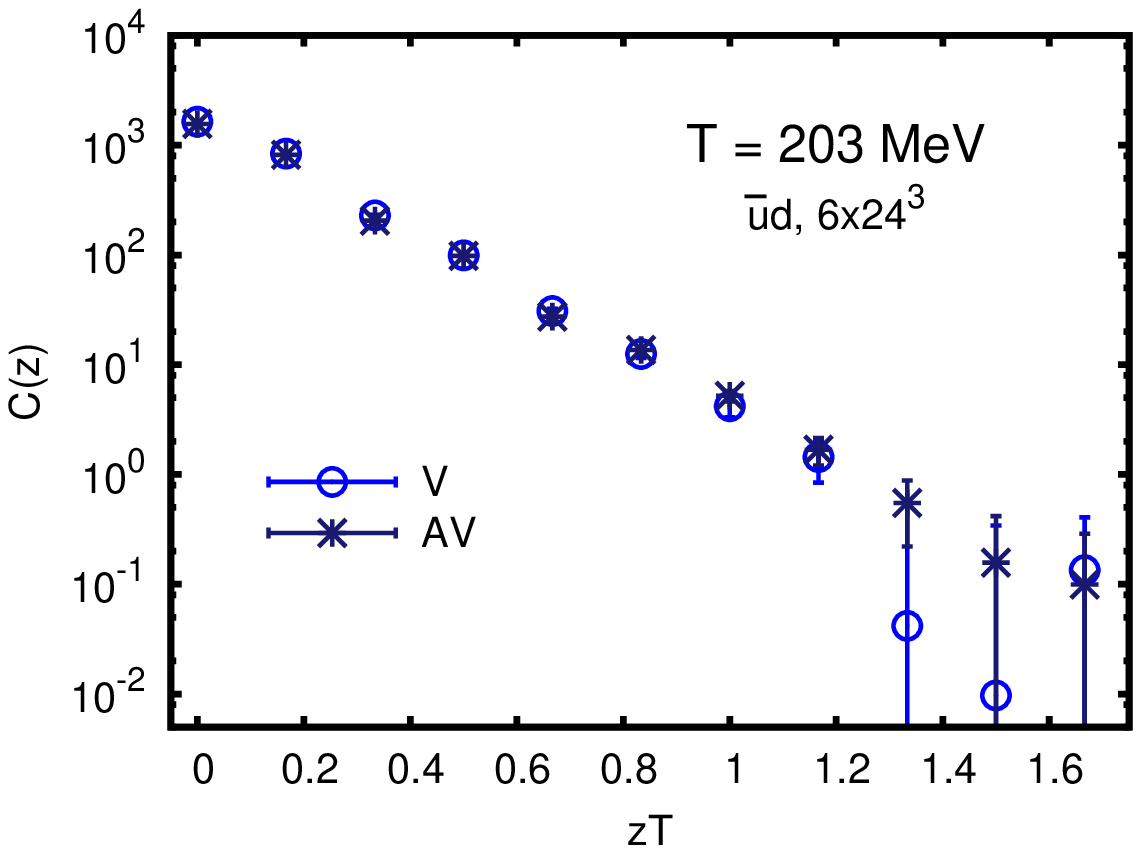}}
\subfigure[]{\label{fig.corr-ps-sc1}\includegraphics[scale=0.50]{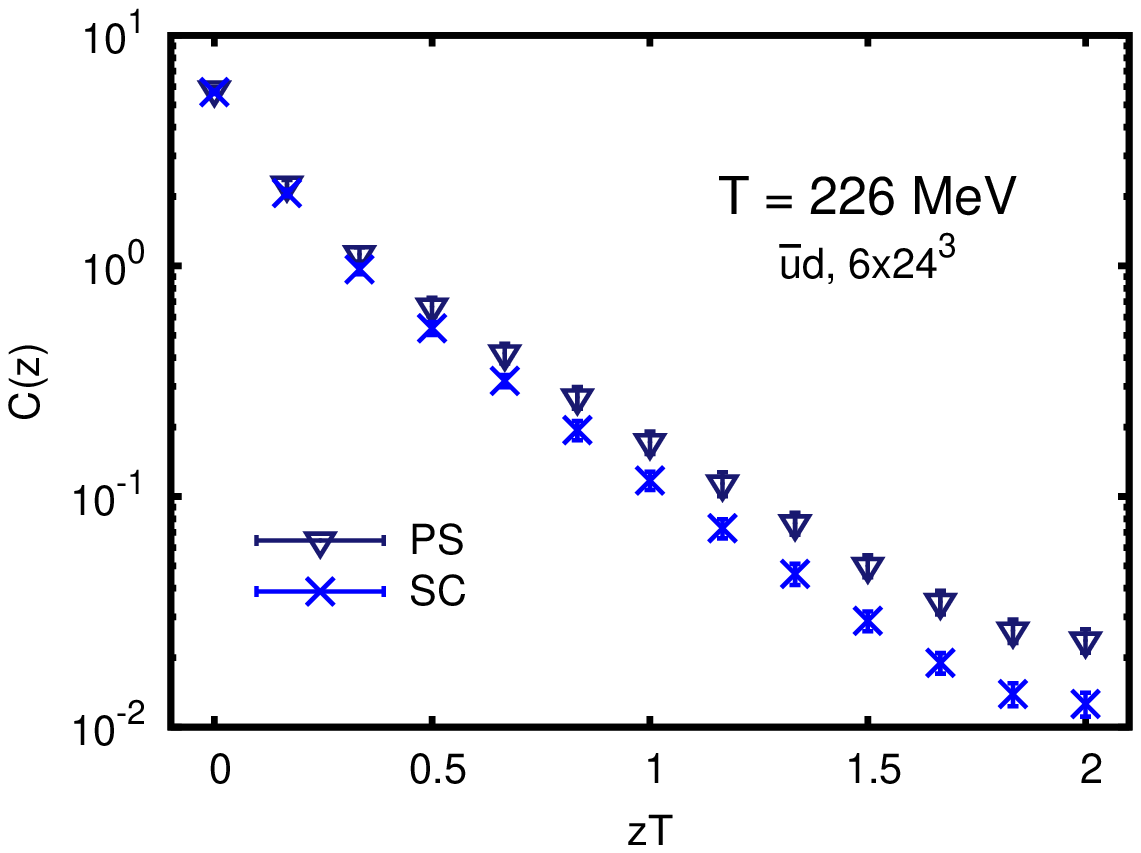}}
\subfigure[]{\label{fig.corr-ps-sc2}\includegraphics[scale=0.50]{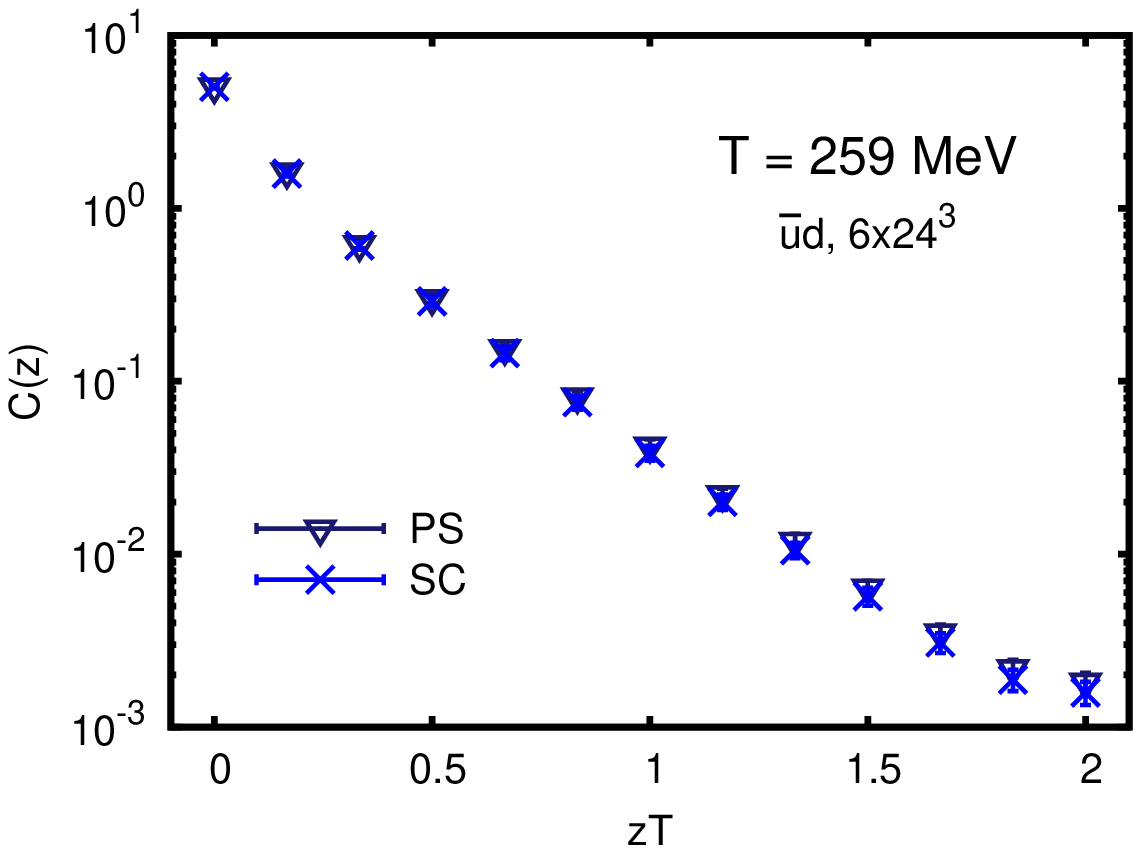}}
\caption{Comparisons of spatial correlation functions for the $\bar ud$ flavor
combination on $N_\tau=6$ lattices: (a) Vector and the axial-vector correlation
functions at $T=203$ MeV using wall sources. The axial-vector correlation function has
been multiplied by the factor $-(-1)^z$ to make it positive for each value of $z$.
Pseudo-scalar and scalar correlation functions at $T=226$ MeV (b) and at $T=259$ MeV
(c) using point sources. The scalar correlation function has been multiplied by
$-(-1)^z$.}
\label{fig.corr}
\end{figure*}

\begin{figure*}[!ht]
\subfigure[]{\label{fig.ps-by-2piT}\includegraphics[scale=0.50]{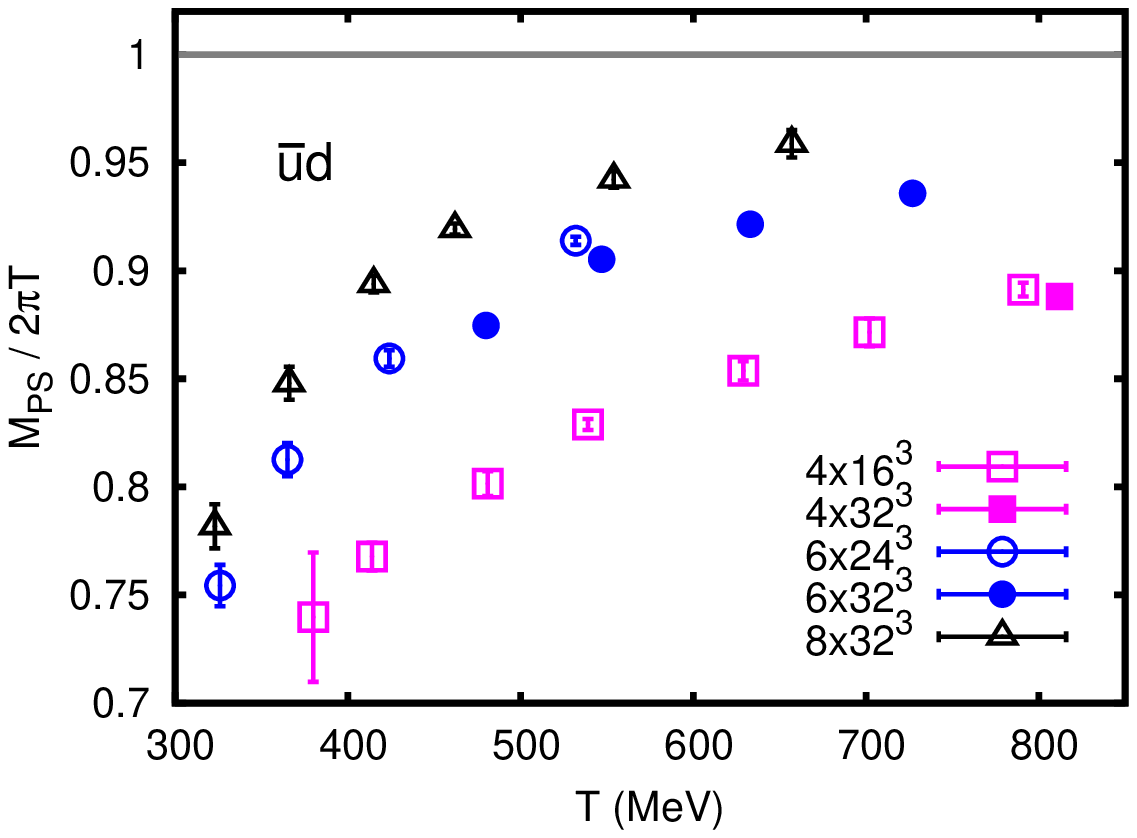}}
\subfigure[]{\label{fig.v-by-2piT}\includegraphics[scale=0.50]{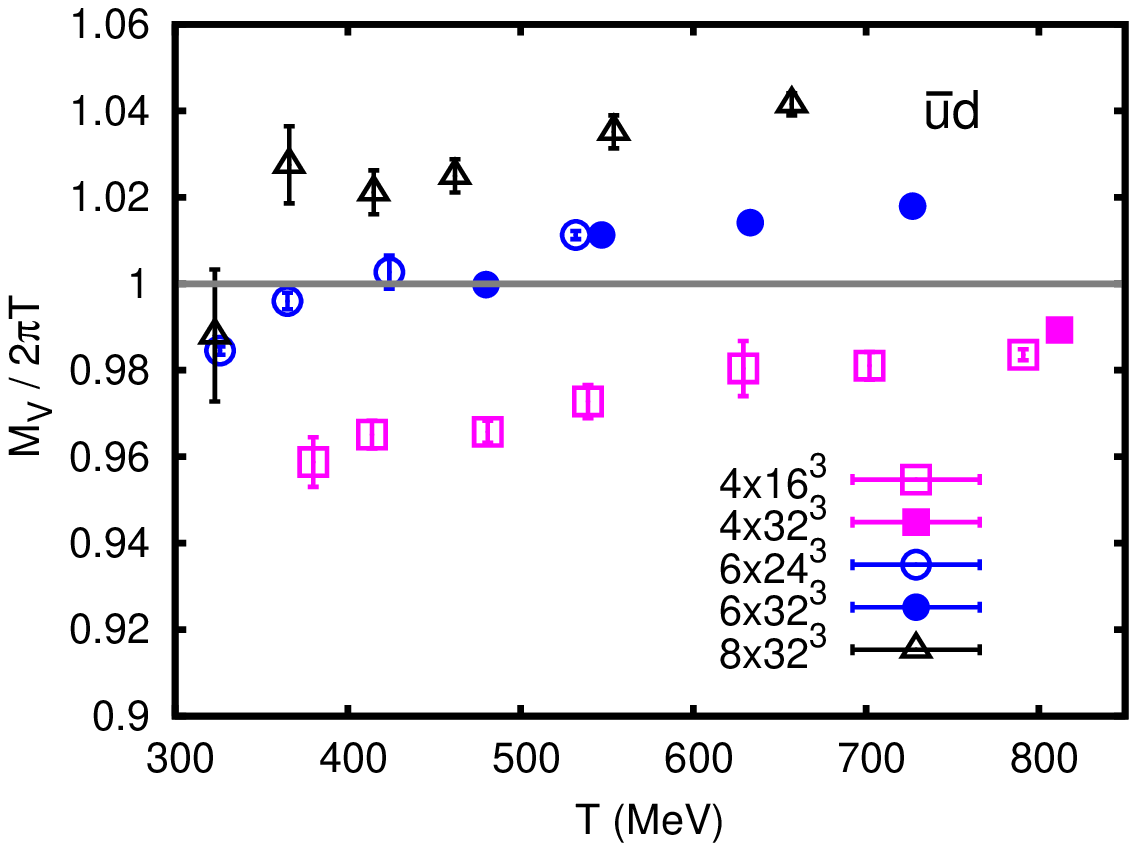}}
\subfigure[]{\label{fig.extrap-ps-v}\includegraphics[scale=0.50]{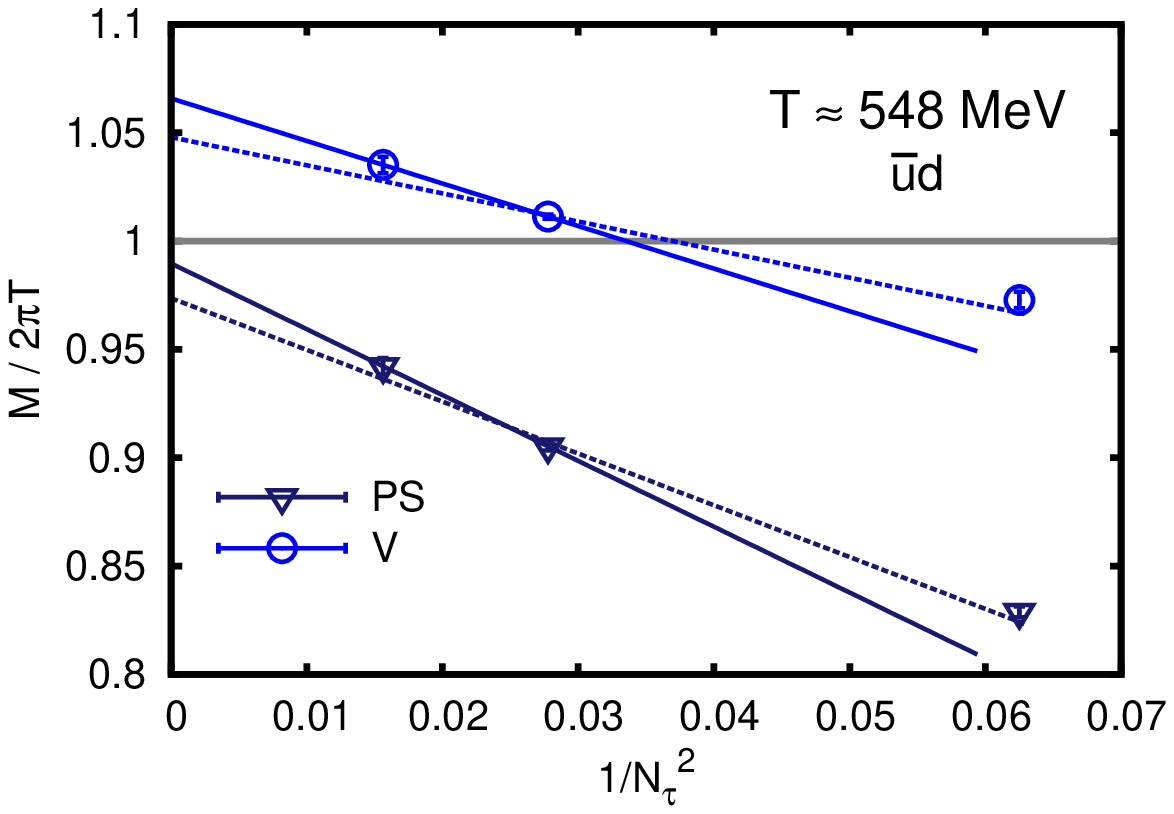}}
\caption{Deviation of the pseudo-scalar (a) and vector (b) screening masses from the
free (continuum) result of $2\pi T$ for the $\bar ud$ sector. (c) Extrapolations,
linear in $1/N_\tau^2$, of the PS and V screening masses to zero lattice spacing at
$T=548(1)$ MeV. The solid lines indicate extrapolations using only $N_\tau=6$
and $8$ data. The dashed lines indicate the extrapolations when $N_\tau=4$ data are
also included in the extrapolations.}
\label{fig.ps-v-highT}
\end{figure*}

In the intermediate temperature regime $T_c\leq T\lesssim1.5T_c$ the spatial
screening correlators show signals of chiral and effective $U_A(1)$ symmetry
restorations. As discussed in Section \ref{se.intro}, chiral symmetry restoration is
reflected through the degeneracy of the V and AV correlation functions at all
distances giving rise to degenerate V and AV screening masses. We found that, for
$N_\tau=6$, $M_V$ becomes approximately equal to $M_{AV}$ at a temperature
$T\simeq200$ MeV. This temperature is approximately equal to the chiral transition
temperature $T_c$ for these lattices identified via the peak of the disconnected
chiral susceptibility. In order to gain some more insight into this degeneracy in
Fig. \ref{fig.corr-v-av} we show the correlation functions of the V and AV channels
at $T=203$ MeV, \ie $T\simeq1.1T_c$, for the $N_\tau=6$ lattice. Indeed we see that
the signature of the chiral symmetry restoration is not only limited to the
degeneracy of the screening masses, which are governed by the long distance behavior
of the correlation functions, but also is reflected in the degeneracy between the
correlation functions at very short distance scales $zT<1$. Similarly, as also
mentioned in Section \ref{se.intro}, the spatial correlation functions of the PS and
the (iso-triplet) SC show effective restoration of $U_A(1)$ for the $2+1$ flavor
theory. In the present work we have found that the PS and SC correlation functions
become degenerate, and hence also the respective screening masses (see Fig.\
\ref{fig.ps-sc_comp}), for $T \gtrsim 1.3T_c$ only.  To illustrate this fact in
detail we show comparisons of the PS and SC screening correlators at $T\simeq1.2T_c$
in Fig.\ \ref{fig.corr-ps-sc1} and at $T\simeq1.3T_c$ in Fig.\ \ref{fig.corr-ps-sc2}.
While at $T\simeq1.2T_c$ the PS and SC screening correlators start differing for
distances $zT\gtrsim0.5$ the correlators become degenerate at all distance scales,
even for $zT<1$, at $T\simeq1.3T_c$. Moreover, as shown in Fig.\
\ref{fig.ps-sc_comp}, the magnitude of the mass splitting $(M_{SC}-M_{PS})$ in the
temperature range $T_c\lesssim T\lesssim1.3T_c$ remains almost unchanged in going
from $\bar ud$ flavor content to $\bar us$. This indicates that in this temperature
regime the mass splitting $(M_{SC}-M_{PS})$, which is sensitive to the presence of
topologically non-trivial configurations \cite{ua1}, remains non-vanishing in the
limit of mass-less light quarks. This conveys the point that the effective $U_A(1)$
restoration in QCD does not happen at the chiral transition temperature $T_c$, \ie
the temperature where the V and AV correlation functions become degenerate. This
observation may have implications in relation to the possible existence of a first
order chiral transition in the limit of two mass-less flavors \cite{nf2-1order}.

As has been pointed out previously in Section \ref{se.intro}, at higher temperatures
$T>1.5T_c$ the mesonic screening masses play an important role in testing the
applicability of the dimensionally reduced perturbation theory. At these temperatures
we have observed that the screening masses in the PS (SC) channel are always lower
than the V (AV) screening masses. In other words, the largest spatial correlation
length belongs to the PS (SC) channel. In order for the dimensional reduction to work
this PS spatial correlation length must be smaller than the smallest spatial
correlation length associated with a glueball like object. As can be seen from Fig.\
\ref{fig.ps-by-2piT}, $M_{PS}/T$ reaches values greater than $4$ at $T\simeq1.5T_c$
and greater than $5$ at $T\simeq2T_c$ for all three lattice spacings. Thus, already
for $T\sim1.5T_c$ the value of the `pion' screening mass is larger than the smallest
glueball screening mass estimated in Ref.\ \cite{GG-glue} and larger than the Debye
mass estimated in Ref.\ \cite{us-debye}. Hence, the lightest fermionic mode is
comparatively heavy already at rather small temperatures such that dimensionally
reduced QCD may work down to temperatures as low as about $1.5 T_c$, as was
demonstrated for a quantity like the spatial string tension
\cite{spatial-string,mikko}. This is in contrast to the previous findings of Ref.\
\cite{GG-glue} where it was concluded that the dimensional reduction framework may
fail for $T\lesssim3T_c$.

In Figs.\ \ref{fig.ps-by-2piT} and \ref{fig.v-by-2piT} we show close-up views of how
the PS and the V screening masses deviate from their continuum free value of $2\pi T$
in the high temperature region $T>1.5T_c$.  In this temperature regime for the
pseudo-scalar (and scalar) channel the ratio $M_{PS}/2\pi T$ not only rises rapidly
towards $1$ with increasing temperature, but also comes closer towards $1$ with
decreasing lattice spacing at a fixed temperature. For example, while on our coarsest
lattice $M_{PS}$ deviates from $2\pi T$ by about $10\%$ at a temperature as high as
$4T_c$ this deviation becomes $\sim5\%$ for our finest lattice already at
$T\simeq3T_c$. On the other hand, the ratio $M_V/2\pi T$ shows a rather slow increase
with increasing temperature. While this ratio remains $\sim2\%$ below $1$ at
$T\simeq4T_c$ for our coarsest lattice spacing it becomes $\sim4\%$ larger than $1$
at $T\simeq3T_c$ for our finest lattice spacing. In view of this, in Fig.\
\ref{fig.extrap-ps-v}, we attempt linear in $a^2\sim1/N_\tau^2$ extrapolations of the
PS and V screening masses at a temperature $T\simeq3T_c$, \ie the maximum temperature
where we have data at all three lattice spacings. The dashed lines in Fig.\
\ref{fig.extrap-ps-v} show extrapolations when all the three lattice spacings are
used during the extrapolations. Such extrapolations give that the PS and the V
screening masses, extrapolated to zero lattice spacing, are respectively 2-3\% below
and 5-6\% above the free theory value of $2\pi T$. As even at this high temperature
the lattice spacing corresponding to $N_\tau=4$ is still quite large (about $0.09$
fm) it is not clear whether including these data in our zero lattice spacing
extrapolations yields reliable results. In order to have some quantitative idea on
the systematics we have also performed extrapolations linear in $1/N_\tau^2$ using
only $N_\tau=6$ and $8$ data. These are indicated by the solid lines in Fig.\
\ref{fig.extrap-ps-v}. Such extrapolations give that the continuum PS screening mass
at this temperature may even be closer to $2\pi T$, smaller only by 1\%, and the
continuum V screening mass is about 7\% larger than $2\pi T$.

\begin{figure}[!t]
\bc
\includegraphics[scale=0.60]{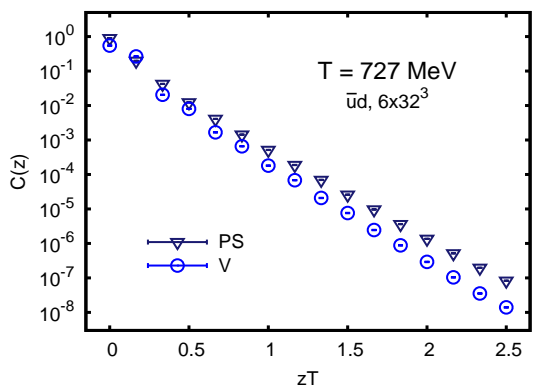}
\ec
\vspace{-0.25cm}
\caption{ A comparison of pseudo-scalar and vector correlation functions, obtained
by using wall sources, on the $N_\tau=6$ lattice at $T=727$ MeV.} 
\label{fig.corr-ps-v}
\end{figure}

Perturbation theory predicts \cite{Mikko,Alberico} that at high temperatures the
screening masses of all the mesons are degenerate and larger than the free theory
value of $2\pi T$. Although we have found that the continuum extrapolated PS
screening mass approaches the continuum free theory value of $2\pi T$ very closely
around $T\simeq3T_c$ and also, for the first time, found evidence that the V
screening masses probably become larger than $2\pi T$ at high temperature, we have
not found any evidence that $M_{PS}$ and $M_{V}$ become degenerate up to temperatures
at least as high as $T\simeq4T_c$. In order to have a closer look at this
non-degeneracy, in Fig.\ \ref{fig.corr-ps-v} we compare directly the PS and V
correlation functions as function of the distance $zT$ around $T\simeq4T_c$. It is
clear from this comparison that the spatial correlators in the PS and V channels
differ from each other at all distances including short ones of $zT<1$. In contrast,
in the (mass-less) free theory the PS and V correlation functions themselves are
degenerate at all distances even on a finite lattice. Although this observation is
subject to cut-off and finite volume effects which could, in principle, be very
different between the free and the interacting theory, we take it as a strong
indication that the non-degeneracy of $M_{PS}$ and $M_{V}$ is probably not a lattice
artifact and may arise due to the presence of spin dependent interactions, as
suggested \eg in \cite{KochShuryak}, at the investigated temperatures. 

As mentioned earlier, one of the sources of systematic errors in our study is the
limited value of the maximum accessible distance $zT$. We have also found moderate
lattice spacing dependence of screening masses at high temperature. Hence, it is
important to corroborate our findings by performing analyses on lattices with larger
aspect ratios and finer lattice spacings. We hope to address this issue in the
future. 

\section*{Acknowledgments}

This work has been supported in part by contracts DE-AC02-98CH10886 and
DE-FG02-92ER40699 with the U.S. Department of Energy, the Bundesministerium f\"ur
Bildung und Forschung under grant 06BI401, the Gesellschaft f\"ur
Schwerionenforschung under grant BILAER, the Helm\-holtz Alliance HA216/EMMI grant
and the Deutsche For\-schungsgemeinschaft under grant GRK 881. The numerical
computations have been carried out on the apeNEXT at Bielefeld University, the QCDOC
computer of the RIKEN-BNL research center, the DOE funded QCDOC at BNL and the
BlueGene/L at the New York Center for Computational Sciences (NYCCS). 

\section*{Appendix}

Tables \ref{tb.nM4}, \ref{tb.nM6} and \ref{tb.nM8} summarize the meson screening
masses for the scalar, pseudo-scalar, averaged transverse axial-vector and averaged
transverse vector channels for the lightest $\bar u d$ flavor combination and for all
the lattices that have been analyzed in this paper. The screening masses are given in
lattice units. The data can be easily converted to either temperature or vacuum
($r_0$) units by means of the included zero temperature results for $r_0/a$ from
\cite{RBCBi-eos}.

\begin{table*}
\begin{center}
\begin{tabular}{|lr|cccc|}
\hline
$\beta$ & $r_0/a$ & $a M_{SC}$ & $a M_{PS}$ & $a M_{AV}$ & $a M_{V}$ \\
\hline
\hline
\multicolumn{6}{|c|}{$16^3 \times 4$} \\
\hline
\multicolumn{6}{|c|}{point sources} \\
\hline
3.290 & 1.823 & 0.860(80) & 0.293(08) &          &          \\
3.320 & 1.908 & 0.702(20) & 0.349(08) &          &          \\
3.351 & 2.069 & 0.695(18) & 0.517(08) & 1.33(12) & 1.37(21) \\
3.382 & 2.230 & 0.802(11) & 0.778(15) & 1.28(10) & 1.33(12) \\
3.410 & 2.503 & 0.920(10) & 0.902(18) & 1.46(06) & 1.46(06) \\
3.460 & 2.890 & 1.046(48) & 1.043(46) & 1.54(09) & 1.52(08) \\
3.540 & 3.687 & 1.162(36) & 1.162(47) & 1.48(05) & 1.49(06) \\
3.570 & 4.009 & 1.206(10) & 1.206(10) & 1.53(01) & 1.53(01) \\
3.630 & 4.651 & 1.259(09) & 1.259(09) & 1.54(01) & 1.54(01) \\
3.690 & 5.201 & 1.306(08) & 1.302(04) & 1.55(01) & 1.55(01) \\
3.760 & 6.050 & 1.341(07) & 1.341(07) & 1.56(02) & 1.57(02) \\
3.820 & 6.752 & 1.369(10) & 1.369(10) & 1.57(02) & 1.57(02) \\
3.920 & 7.590 & 1.400(05) & 1.400(05) & 1.56(01) & 1.56(01) \\
\hline
\multicolumn{6}{|c|}{wall sources} \\
\hline
3.382 & 2.230 & 0.801(13) & 0.762(19) & 1.380(60) & 1.380(50) \\
3.410 & 2.503 & 0.905(11) & 0.890(13) & 1.460(60) & 1.450(70) \\
3.460 & 2.890 & 1.054(22) & 1.051(22) & 1.515(35) & 1.520(38) \\
3.540 & 3.687 & 1.166(10) & 1.166(10) & 1.506(09) & 1.506(09) \\
3.570 & 4.009 & 1.204(06) & 1.204(06) & 1.516(05) & 1.516(05) \\
3.630 & 4.651 & 1.255(05) & 1.255(05) & 1.517(04) & 1.517(04) \\
3.690 & 5.201 & 1.291(05) & 1.291(05) & 1.528(06) & 1.528(06) \\
3.760 & 6.050 & 1.330(04) & 1.330(04) & 1.540(10) & 1.540(10) \\
3.820 & 6.752 & 1.355(04) & 1.355(04) & 1.541(05) & 1.541(05) \\
3.920 & 7.590 & 1.386(02) & 1.386(02) & 1.545(02) & 1.545(02) \\
\hline\hline
\multicolumn{6}{|c|}{$32^3 \times 4$} \\
\hline
\multicolumn{6}{|c|}{point sources} \\
\hline
3.920 & 7.814 &           & 1.395(2) &           & 1.560(1) \\
\hline
\multicolumn{6}{|c|}{wall sources} \\
\hline
3.920 & 7.814 &           & 1.392(1) &           & 1.554(2) \\
\hline
\end{tabular}
\end{center}
\caption{Screening masses from $N_\tau = 4$ lattices.
}
\label{tb.nM4}
\end{table*}

\begin{table*}
\begin{center}
\begin{tabular}{|lr|cccc|}
\hline
$\beta$ & $r_0/a$ & $a M_{SC}$ & $a M_{PS}$ & $a M_{AV}$ & $a M_{V}$ \\
\hline
\hline
\multicolumn{6}{|c|}{$24^3 \times 6$} \\
\hline
\multicolumn{6}{|c|}{point sources} \\
\hline
3.351 &  2.069 & 0.900(30) & 0.257(01) &         &         \\
3.410 &  2.503 & 0.700(20) & 0.226(02) &         &         \\
3.430 &  2.647 & 0.615(09) & 0.216(03) &         &         \\
3.445 &  2.813 & 0.510(20) & 0.253(05) &         &         \\
3.455 &  2.856 & 0.420(10) & 0.292(07) &         &         \\
3.460 &  2.890 & 0.435(08) & 0.334(08) &         &         \\
3.490 &  3.223 & 0.490(10) & 0.440(10) &         &         \\
3.510 &  3.426 & 0.580(10) & 0.560(10) &         &         \\
3.540 &  3.687 & 0.648(09) & 0.640(08) & 1.11(6) & 1.08(6) \\
3.570 &  4.009 & 0.728(08) & 0.726(08) & 1.04(3) & 1.04(3) \\
3.630 &  4.651 & 0.790(10) & 0.790(10) & 1.04(1) & 1.04(2) \\
3.690 &  5.201 & 0.851(08) & 0.851(08) & 1.04(1) & 1.04(1) \\
3.760 &  6.050 & 0.900(04) & 0.900(04) & 1.06(1) & 1.06(1) \\
3.920 &  7.590 & 0.957(02) & 0.957(02) & 1.07(1) & 1.07(1) \\
\hline
\multicolumn{6}{|c|}{wall sources} \\
\hline
3.351 &  2.069 & 1.070(50) & 0.270(3)  & 1.650(20) & 1.270(30) \\
3.410 &  2.503 & 0.780(20) & 0.240(2)  & 1.408(09) & 1.125(05) \\
3.430 &  2.647 & 0.685(02) & 0.227(1)  & 1.140(50) & 0.985(06) \\
3.445 &  2.813 & 0.512(02) & 0.246(1)  & 1.090(04) & 0.994(04) \\
3.455 &  2.856 & 0.456(02) & 0.262(2)  & 1.058(05) & 0.990(10) \\
3.460 &  2.890 & 0.436(01) & 0.316(1)  & 1.010(08) & 1.030(10) \\
3.490 &  3.223 & 0.487(01) & 0.438(1)  & 0.990(06) & 0.990(07) \\
3.510 &  3.426 & 0.562(01) & 0.538(2)  & 1.009(04) & 1.002(04) \\
3.570 &  4.009 & 0.705(01) & 0.702(1)  & 1.016(01) & 1.011(01) \\
3.630 &  4.651 & 0.788(01) & 0.788(1)  & 1.031(01) & 1.031(01) \\
3.690 &  5.201 & 0.839(01) & 0.839(1)  & 1.043(02) & 1.043(02) \\
3.760 &  6.050 &           &           & 1.049(05) & 1.050(04) \\
3.920 &  7.590 & 0.938(01) & 0.941(1)  & 1.059(01) & 1.059(01) \\
\hline\hline
\multicolumn{6}{|c|}{$32^3 \times 6$} \\
\hline
\multicolumn{6}{|c|}{point sources} \\
\hline
3.820 &  6.864 &           & 0.916(3) &         & 1.059(4) \\
3.920 &  7.814 &           & 0.948(2) &         & 1.061(4) \\
4.000 &  9.048 &           & 0.965(3) &         & 1.063(4) \\
4.080 & 10.390 &           & 0.980(3) &         & 1.064(2) \\
\hline
\multicolumn{6}{|c|}{wall sources} \\
\hline
3.820 &  6.864 &           & 0.911(2) &         & 1.047(2) \\
3.920 &  7.814 &           & 0.946(1) &         & 1.059(1) \\
4.000 &  9.048 &           & 0.961(1) &         & 1.062(2) \\
4.080 & 10.390 &           & 0.978(2) &         & 1.066(2) \\
\hline
\end{tabular}
\end{center}
\caption{Screening masses from $N_\tau = 6$ lattices.
}
\label{tb.nM6}
\end{table*}

\begin{table*}
\begin{center}
\begin{tabular}{|lr|cccc|}
\hline
$\beta$ & $r_0/a$ & $a M_{SC}$ & $a M_{PS}$ & $a M_{AV}$ & $a M_{V}$ \\
\hline
\hline
\multicolumn{6}{|c|}{$32^3 \times 8$} \\
\hline
\multicolumn{6}{|c|}{point sources} \\
\hline
3.430 &  2.647 &            & 0.201(01) &           &           \\
3.500 &  3.328 & 0.463(174) & 0.175(01) &           &           \\
3.570 &  4.009 & 0.366(010) & 0.311(14) &           &           \\
3.585 &  4.160 & 0.382(010) & 0.351(10) &           &           \\
3.760 &  6.050 & 0.615(008) & 0.614(08) & 0.777(12) & 0.776(12) \\
3.820 &  6.864 & 0.666(006) & 0.666(06) & 0.807(07) & 0.807(07) \\
3.920 &  7.814 & 0.702(003) & 0.702(03) & 0.802(04) & 0.802(04) \\
4.000 &  9.048 & 0.721(001) & 0.722(02) & 0.803(02) & 0.805(03) \\
4.080 & 10.390 & 0.740(003) & 0.740(03) & 0.813(03) & 0.813(03) \\
4.200 & 12.420 & 0.753(006) & 0.753(05) & 0.831(12) & 0.816(09) \\
\hline
\multicolumn{6}{|c|}{wall sources} \\
\hline
4.000 &        & 0.716(001)   & 0.716(03)  & 0.799(04) & 0.800(03)  \\
4.200 &        & 0.755(002)   & 0.755(01)  & 0.832(12) & 0.818(02)  \\
\hline
\end{tabular}
\end{center}
\caption{Screening masses from $N_\tau = 8$ lattices.
}
\label{tb.nM8}
\end{table*}


\end{document}